%% file: ms.tex
\documentclass{aa}

\usepackage{graphicx}
\usepackage[varg]{txfonts}

\newcommand{\feh}{\mathrm{[Fe/H]}}

\newcommand{\xfe}{\mathrm{[X/Fe]}}
\newcommand{\teff}{T_\mathrm{eff}}
\newcommand{\logg}{\log g}
\newcommand{\afe}{A_\mathrm{Fe}}

\newcommand{\fei}{Fe\,\textsc{i}}
\newcommand{\feii}{Fe\,\textsc{ii}}

\newcommand{\kms}{km\,s$^{-1}$}
\newcommand{\tc}{T_\mathrm{C}}

\begin{document} 

\title{Chemical signatures of planets: beyond solar-twins\thanks{
       Based on observations collected at the 
       European Organisation for Astronomical Research 
       in the Southern Hemisphere, Chile, observing proposals
       086.D0062 and 087.D0010.}
      }

\author{I. Ram\'irez\inst{1}\fnmsep\thanks{NASA Sagan Fellow.} \and
        J. Mel\'endez\inst{2} \and
        M. Asplund\inst{3}
       }

\institute{Department of Astronomy, University of Texas at Austin;
           2515 Speedway, Stop C1400, Austin, TX 78712-1205, USA;\\
           \email{ivan@astro.as.utexas.edu}
           \and
           Departamento de Astronomia do IAG/USP,
           Universidade de S\~ao Paulo;
           Rua do M\~atao 1226, S\~ao Paulo, 05508-900, SP, Brasil;\\
           \email{jorge.melendez@iag.usp.br}
           \and
           Research School of Astronomy and Astrophysics,
           The Australian National University;
           Cotter Road, Weston, ACT 2611, Australia;\\
           \email{martin.asplund@anu.edu.au}
          }

\date{Received --- --, ---; accepted --- --, ---}

\abstract
{Elemental abundance studies of solar twin stars suggest that the solar chemical composition contains signatures of the formation of terrestrial planets in the solar system, namely small but significant depletions of the refractory elements.}
{To test whether these chemical signatures of planets are real, we study stars which, compared to solar twins, have less massive convective envelopes (therefore increasing the amplitude of the predicted effect) or are, arguably, more likely to host planets (thus increasing the frequency of signature detections).}
{We measure relative atmospheric parameters and elemental abundances of two groups of stars: a ``warm'' late-F type dwarf sample (52 stars), and a sample of ``metal-rich'' solar analogs (59 stars). The strict differential approach that we adopt allows us to determine with high precision ($\mathrm{errors}\sim0.01$\,dex) the degree of refractory element depletion in our stars independently of Galactic chemical evolution. By examining relative abundance ratio versus condensation temperature plots we are able to identify stars with ``pristine'' composition in each sample and to determine the degree of refractory-element depletion for the rest of our stars. We calculate what mixture of Earth-like and meteorite-like material corresponds to these depletions.}
{We detect refractory-element depletions with amplitudes up to about 0.15\,dex. The distribution of depletion amplitudes for stars known to host gas giant planets is not different from that of the rest of stars. The maximum amplitude of depletion increases with effective temperature from 5650\,K to 5950\,K, while it appears to be constant for warmer stars (up to 6300\,K). The depletions observed in solar twin stars have a maximum amplitude that is very similar to that seen here for both of our samples.}
{Gas giant planet formation alone cannot explain the observed distributions of refractory-element depletions, leaving the formation of rocky material as a more likely explanation of our observations. More rocky material is necessary to explain the data of solar twins than metal-rich stars, and less for warm stars. However, the sizes of the stars' convective envelopes at the time of planet formation could be regulating these amplitudes. Our results could be explained if disk lifetimes were shorter in more massive stars, as independent observations indeed seem to suggest. Nevertheless, to reach stronger conclusions we will need a detailed knowledge of extrasolar planetary systems down to at least one Earth mass around a significant number of stars.}


\maketitle

\section{Introduction}

Based on our knowledge of the solar system, we expect the chemical composition of planets, particularly those which are Earth-like, to be different from that observed in the photospheres of their host stars. Since they formed essentially at the same time and from the same gas cloud, it is reasonable to propose that the process of planet formation leaves chemical signatures on the planet-host stars. In summary, the photospheres of stars that host planets are expected to be deficient in elements which are abundant in planets compared to stars that did not form them. This is because those missing elements were left behind in the planets and other smaller objects (e.g., asteroids) that formed around that time. In practice, this picture is complicated by the fact that the amount of metals taken away from the star by the planets may be too small to be detected by current observational means.

\citet[hereafter M09]{melendez09:twins} found that, compared to a sample of 11 so-called solar twin stars -- objects with spectra nearly indistinguishable from the solar one -- the Sun is deficient in refractory elements relative to volatiles. Assuming that the volatile element abundance is normal, the amount of refractory element depletion observed in the Sun is compatible with the total mass of rock formed in the solar system \cite[e.g.,][]{chambers10,melendez12}. M09 argue that this peculiar solar chemical composition is the end result of the formation of terrestrial planets and other rocky bodies in the solar system. A natural implication of this hypothesis is that most other solar twins did not form as many rocky bodies, which seems arbitrary. Indeed, recent results from the {\it Kepler} Mission suggest that planetary systems with rocky planets with total mass greater than that of the solar system rocks may be as common as the solar case \citep{fressin13}. However, the internal composition of super-Earths is highly uncertain. Moreover, it is also possible that the amplitude of refractory element depletions is regulated by other early stellar evolution processes, as explained below.

In related work, \citet[hereafter R11]{ramirez11} showed that the overall metallicity of the secondary star in the 16\,Cygni binary system, which is known to be a gas giant planet host \citep{cochran97}, is slightly lower than that of the primary, which does not have a planet detected yet. They proposed that the elemental abundance difference between the two 16\,Cygni stars was created when the giant planet around 16\,Cygni\,B formed. In this case, both volatile and refractory elements are equally depleted around the planet-host component.

When examining M09's and R11's hypotheses, one must take into account the fact that in order for the planet signature to be imprinted, stars' convective envelopes are required to have a low mass at the time of planet formation. More precisely, they need to be small when planetesimals form. This is not in agreement with classical models of stellar interiors and evolution, which suggest that stars like the Sun are born fully convective \cite[e.g.,][]{iben65}. The radiative zone is developed in about 10-30\,Myr, gradually shrinking the convective envelope \cite[e.g.,][]{serenelli11}. Planetesimals are expected to form within the first 10\,Myr of the star's life \cite[e.g.,][]{fedele10}, i.e., at a time when the star's convective envelope is still massive.

Star formation with episodic accretion, which is supported by both theory and observations \cite[e.g.,][]{enoch09,vorobyov09,dunham10,kim11}, provides one way of solving the problem described above. Contrary to the classical scenario, in these models stars are formed with variable accretion rate. The stars' interiors heat up quicker, developing radiative cores and thin convective envelopes faster than their classical counterparts \citep{baraffe10}. For some accretion histories, a thin convective envelope can be formed as quickly as in 5\,Myr, allowing at least in principle to imprint the chemical planet signatures. Moreover, this implies that the particular episodic accretion history of a star that forms terrestrial planets determines whether the signature is imprinted or not.

Other works have examined detailed chemical abundances in Sun-like stars and their possible connection to exoplanets \cite[e.g.,][]{ramirez09,ramirez10,gonzalez10,gonzalez-hernandez10,gonzalez-hernandez13,schuler11:16cyg,schuler11}. Small chemical element depletions are generally detected in these other works, but their interpretations may be different. Certainly, there are caveats in the M09 and R11 interpretations as well as independent observations that appear to contradict their results. Nevertheless, the idea of planet formation imprinting signatures on stellar chemical abundances is very appealing, if confirmed. 
Chemical abundance analysis of stars is a straightforward process. 
The prospect of being able to use relatively simple photospheric chemical analysis to find or confirm the presence of both terrestrial and gas giant planets around distant stars highlights the importance of investigating in detail the M09 and R11 hypothesis.

The M09 and R11 works employed solar twin and analog stars. The sizes of the convective envelopes of all objects analyzed by them are very similar, which helped the interpretation of their findings. In this work, we investigate the proposed chemical signatures of planet formation using two samples of stars which are 1) warmer and 2) more metal-rich than the Sun, in order to determine whether they follow the expected behavior given their different convective envelope sizes \cite[e.g.,][]{pinsonneault01}. In addition, the high-metallicity sample could help investigating the impact of a higher frequency of planets on these signatures, because of the well-known planet-metallicity correlation \cite[e.g.,][]{gonzalez98,santos04,fischer05}. We acknowledge, however, that this correlation seems to be valid only for gas giant planets \cite[e.g.,][]{mayor11,buchhave12}.

\section{Data and Spectroscopic Analysis}

\subsection{Sample selection and observations} \label{s:sample}

\begin{figure}
\includegraphics[bb=75 370 390 742,width=9.3cm]
{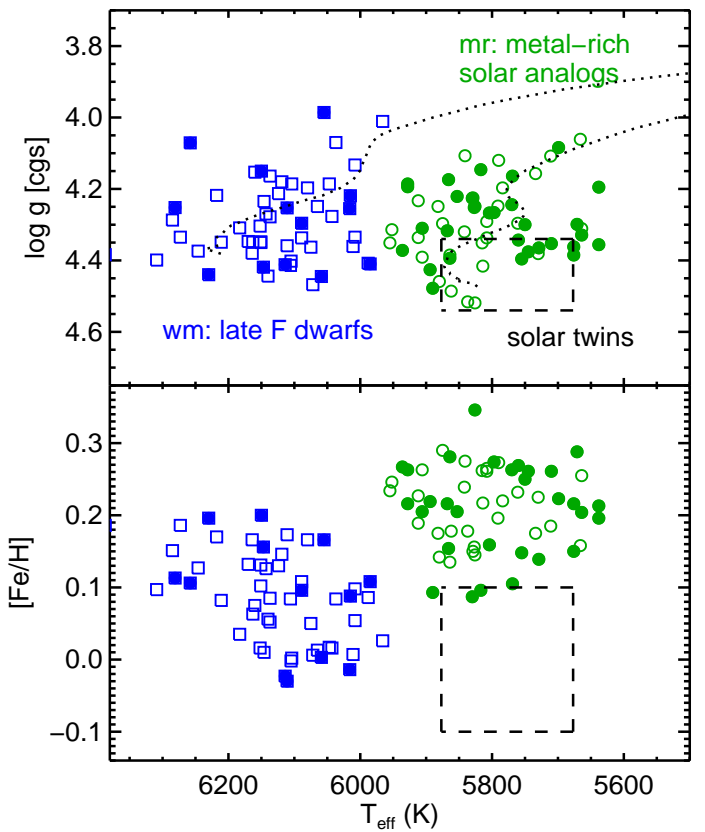}
\caption{Distribution of atmospheric parameters for our warm (squares) and metal-rich (circles) samples. Filled symbols represent known planet hosts. The rectangles show the regions occupied by solar twin stars. In the top panel, Yonsei-Yale evolutionary tracks for $M=1.2\,M_\odot$, $\feh=+0.1$ (representative of our wm sample) and $M=1.1\,M_\odot$, $\feh=+0.2$ (a typical mr star) are shown with dotted lines.}
\label{f:sample}
\end{figure}

Two samples of stars were constructed for this work: a ``warm'' (wm) F-dwarf sample and a ``metal-rich'' (mr) solar analog sample. We employed a large database of previously published stellar atmospheric parameters to search for these stars and observed 52 (59) wm (mr) stars.\footnote{This catalog is maintained by J.\,Mel\'endez and it is similar to, but more comprehensive than other available compilations such as those by \cite{cayrel01} and \cite{soubiran10}.} The distribution of these stars on the $\teff$ (effective temperature) vs.\ $\logg$ (logarithmic surface gravity) and $\teff$ vs.\ $\feh$ (iron abundance) planes is shown in Figure~\ref{f:sample}.\footnote{Here we use the standard notation for elemental abundances: $A_\mathrm{X}=\log(n_\mathrm{X}/n_\mathrm{H})+12$, where $n_\mathrm{X}$ is the number density of element X, and $\mathrm{[X/H]}=A_\mathrm{X}-A_\mathrm{X}^\odot$.} The mean $\teff$ of the wm sample was chosen so that the fraction of stars with high projected rotational velocity ($V\sin i$) is relatively small, yet hot enough that the sizes of these stars' convective envelopes are significantly smaller than those of solar twin stars. Low $V\sin i$ values minimize the impact of line blending due to rotational broadening, allowing us to measure single line strengths with high accuracy. However, this choice naturally biases our sample towards inactive stars. For the mr sample we forced a mean $\teff$ equal to solar. The metallicities of both samples are super-solar, but more so for the mr sample. This choice was made deliberately to include an important number of known planet hosts in both groups.

We employed the exoplanets.org online database \citep{wright11} to identify the stars from our samples which are known to host planets (filled symbols in Figure~\ref{f:sample}) and to assign planet properties such as minimum mass. In the case of multi-planet systems we adopted the minimum mass of the more massive planet. There are 13 known planet hosts in the wm sample and 30 in the mr sample, which corresponds to 25\,\% and 51\,\% of the total number of stars in each sample. Since not all stars in our samples have been searched for planets, the open symbols in Figure~\ref{f:sample} do not necessarily represent non-planet-hosts. In fact, even for most of those stars that have been searched for planets and none have been found yet, only short-period gas giants can be really excluded.

Spectroscopic observations of our stars were carried out with the Ultraviolet and Visual Echelle Spectrograph \cite[UVES,][]{dekker00} on the Unit Telescope 2 (UT2) of the Very Large Telescope (VLT) array, operating in service mode during the European Southern Observatory (ESO) observing periods 86 (September 2010 to March 2011) and 87 (April to August 2011). We employed the $0.7$\,arcsec and $0.3$\,arcsec slits for the blue and red arms, which deliver spectral resolutions ($R=\lambda/\Delta\lambda$) of 65,000 and 110,000, respectively. We used the DIC2 (dichroic) 390/760\,nm standard setting, which results in spectral coverage from 326 to 445\,nm in the blue arm and from 565 to 946\,nm in the red arm. Exposure times were set so that a similar signal-to-noise ratio was achieved for all objects ($S/N\simeq400$ at 650\,nm); they ranged between 1 and 45 minutes. We reduced our UVES spectra in the standard manner using IRAF's echelle package.\footnote{IRAF is distributed by the National Optical Astronomy Observatory, which is operated by the Association of Universities for Research in Astronomy (AURA) under cooperative agreement with the National Science Foundation.}

\begin{table}
\caption{Sample\tablefootmark{a}}
\centering
\tiny
\begin{tabular}{lcccrrr}\hline\hline
HIP & $V_\mathrm{mag}$ & $\teff$ & $\logg$ & $\feh$ & $N_\mathrm{lit}$ & $M_\mathrm{planet}$\tablefootmark{b} \\ 
& & (K) & [cgs] & & & ($M_\mathrm{Jupiter}$) \\ \hline
\input{sample_small.tex}\hline
\end{tabular}
\tablefoot{\tiny
\tablefoottext{a}{The atmospheric parameters listed here correspond to those from our literature compilation. Column $N_\mathrm{lit}$ shows the number of published values.}
\tablefoottext{b}{Minimum mass of known planet hosted or that of the most massive planet for known multi-planet system hosts.}
}
\label{t:sample}
\end{table}

Table~\ref{t:sample} lists our sample stars along with their atmospheric parameters from the literature and other relevant information.

\subsection{Atmospheric parameters} \label{s:atmospheric_parameters}

To determine the stars' fundamental atmospheric parameters $\teff,\logg,\feh$ we employed a differential iron line analysis. Iron abundances were measured using the 2010 version of the spectrum synthesis code MOOG,\footnote{http://www.as.utexas.edu/$\sim$chris/moog.html} employing the Kurucz ``odfnew'' model atmosphere grid.\footnote{http://kurucz.harvard.edu/grids.html} The linelist adopted was constructed from the one used in \cite{ramirez13:thin-thick}.\footnote{This line list contains only lines with strengths that are low enough to be on the linear part of the curve of growth in a typical solar analysis, thus reducing the impact of saturation in the determination of elemental abundances of solar-type stars.} We inspected each of the lines listed in that work and kept only those that appeared clean (i.e., unblended) and fell in a spectral region with high local $S/N$. Equivalent widths ($EW$s) were measured using IRAF's splot tool. Each line was first inspected in all spectra to determine the approximate location of continuum windows, which were then applied consistently to all objects. Gaussian profiles were fit to each line to determine the $EW$ values.

Since our goal is to achieve the highest precision possible in relative abundances, instead of using a solar spectrum as reference in our spectroscopic analysis, we performed star-to-star differential analyses within each sample. This approach minimizes the impact of systematic errors in the same way that the analysis of solar twin stars using the solar spectrum as reference did in the M09 work.\footnote{A similar approach has been independently taken in the analysis of giant stars in NGC 6752 by \cite{yong13}, where giant stars with similar stellar parameters were analyzed differentially using as reference a giant with parameters near the mean value of the sample, achieving thus uncertainties as low as $\sim$0.01 dex.} As shown by R11, even in the case of the solar analog stars of the 16\,Cygni binary system, a direct comparison of the two component stars results in higher precision compared to the case in which abundances relative to solar are first measured and then used to find the differences between 16\,Cygni A and B. In this work, we go one step further and determine relative parameters and abundances of every star relative to each of the other ones within their sample (warm or metal-rich), and use all the available information to reduce the observational errors.

Our procedure is as follows. For each pair of spectra ($i,j$), relative atmospheric parameters were determined, for example $\Delta\teff(i,j)$. The procedure to derive each of these relative parameters is standard: we modified them iteratively until no correlations of the relative iron abundance with either excitation potential or reduced equivalent width were present. Also, they were set so that the mean iron abundances inferred from \fei\ and \feii\ lines agree. All three relative parameters $\Delta\teff$, $\Delta\logg$, and $\Delta v_t$ were modified simultaneously in each iteration. The relative iron abundances $\Delta\feh$ were measured using a line-by-line differential approach, which minimizes the impact of errors in the atomic transition probabilities and possibly also other systematics.\footnote{The microturbulent velocity $v_t$ is determined essentially by minimizing the correlation between iron abundance and $EW$. In this work we use the $EW$ values measured in our spectra, but note that \cite{magain84} suggests to employ expected $EW$ values instead, in order to prevent an overestimate of the $v_t$ values.}

The end result from the calculations described above is a $n\times n$ matrix for each parameter, where $n$ is the number of stars in each sample. Due to observational errors, these matrices are not symmetric. Although in most cases we find that $\Delta\teff(i,j)$ and $\Delta\teff(j,i)$ are exactly the same, for noisy data it is not rare to find that $\Delta\teff(i,j)\neq\Delta\teff(j,i)$. More importantly, in general we find that $\Delta\teff(i,j)\neq\Delta\teff(i,k)+\Delta\teff(k,j)$, contrary to what is expected in an idealized situation (i.e., when observational and systematic errors are equal to zero). Under the real conditions, one is faced with the question of what set of relative parameters to adopt (i.e., which row or column from the $n\times n$ matrix). To solve this problem we used the ``self-improvement'' technique described in \cite{allende07}, which forces a unique, consistent solution for any matrix of relative values. This procedure takes as input the observational matrix (and its associated error matrix) and uses all the available data to force $\Delta\teff(i,j)=\Delta\teff(j,i)$ and $\Delta\teff(i,j)=\Delta\teff(i,k)+\Delta\teff(k,j)$. In principle, self-improvement also reduces the intrinsic, observational errors, but our data is of such high quality that this reduction of internal error was minor and not really noticeable. However, self-improvement does ensure that the impact of outliers within the matrix is minimized because they end up being ``absorbed'' by the good data points when calculating the final result using all other elements of the relative abundance matrix.

A natural concern of our strict differential approach is the initial guess values for the star's atmospheric parameters. These numbers are used as reference in each of the computations described above. We employed as guess values those from the literature compilation mentioned in Section~\ref{s:sample}. Also, keeping track of the errors in this scheme is not straightforward. To investigate the impact of inaccurate input values on the relative parameters, and to estimate our internal errors, we ran a simulation as follows.

First, we created distributions of stars in stellar parameter space with the same mean and standard deviation values as our actual warm and metal-rich samples, assuming Gaussian distributions with no underlying stellar population. We adopted the parameters from this created distribution as the ``real'' parameters in this simulation. Next, we employed the ewfind driver of MOOG to compute equivalent widths for all the iron lines employed in this work using the real parameters for each simulated star. We introduced a Gaussian error to these $EW$ values with standard deviations of 2.5\,\% for the warm sample and 1.8\,\% for the metal-rich sample. The justification to choose these numbers is that the internal errors in the derived parameters which result with them (within the simulation and for a given pair of spectra) are consistent with those obtained using the actual data. $EW$ errors of $\sim2$\,\% are reasonable for data of high quality as ours, and it is expected that stars in our warm sample have larger $EW$ errors due to their higher $V\sin i$, which favors line blending by weak features.

Then, we created a table of guess parameters, using the real ones as starting point, and introducing random Gaussian variations of 1-$\sigma=70$\,K in $\teff$, 0.06\,dex in $\logg$, and 0.05\,dex in $\feh$. Although these 1-$\sigma$ values appear too optimistic, we note that most of our stars are well-studied and the literature compilation contains several entries which have been averaged for our work. Thus, the guess parameters of our sample stars are in fact reasonably well constrained. Indeed, comparison of these literature parameters to our finally derived values shows mean differences that have 1-$\sigma$ scatter values of 52 and 31\,K in $\teff$, 0.08 and 0.07\,dex in $\logg$, and 0.04 and 0.03 in $\feh$ for our warm and metal-rich samples, respectively.

Finally, we ran our codes for stellar parameter determination using the guess parameters as input. Our final parameter solutions were then compared to the real values of the simulation. We find that the relative $\teff$ values (e.g., $\Delta\teff(i,j)$) are recovered with a 1-$\sigma$ error of 14\,K for the warm sample and 13\,K for the metal-rich sample. Corresponding values for $\logg$ are 0.03\,dex for the warm sample and 0.02\,dex for the metal-rich samples, while those for $\feh$ are 0.012\,dex in both cases.

Obviously the absolute parameters $\teff,\logg,\feh$ are only as good as those of the one star that we decide to pick as reference. The choice of reference is trivial in the case of solar twins where one must use the Sun, whose canonical values of $\teff,\logg,\feh$ have zero error. In our case we chose as references the two stars with the largest number of published stellar parameters (see next Section).

M09 achieved 0.01\,dex precision in [Fe/H] using data of similar quality as ours. The fact that we obtain a comparable precision suggests that our approach is reliable and will produce results that can be interpreted in a similar way as in the M09 work.

\subsection{Standard stars and elemental abundances}

In order to derive abundances of other elements, relative or absolute, it is necessary to adopt absolute stellar parameters for our sample stars. This can be achieved by using the relative parameters derived as described in the previous Section and adopting the absolute parameters of a given reference star. The derived abundances (both relative and absolute) will be dependent on the choice of reference star, but we expect this dependency to be less important if we pick a representative, well-studied star from each sample. Thus, we searched for stars that have an important number of published parameters and are known to be well-behaved (i.e., non-variable spectra showing low levels of activity and no evidence of binarity). The two stars chosen as reference in our work are HIP\,14954 (for the warm sample) and HIP\,74500 (for the metal-rich sample).

\begin{figure}
\includegraphics[bb=75 372 390 572,width=8.5cm]
{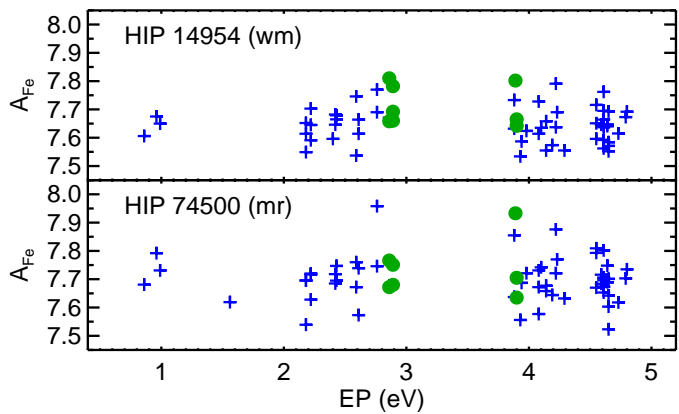}
\caption{Absolute iron abundance of our two reference stars for each of the lines employed in this work as a function of excitation potential. Crosses represent \fei\ lines and circles correspond to \feii\ lines.}
\label{f:reference}
\end{figure}

HIP\,14954 has 27 entries in our literature compilation while HIP\,74500 has 15. We adopted the slightly rounded-off robust mean (trimean) of these literature values (Table \ref{t:reference}). We computed iron abundances using MOOG, not in a strict differential manner, but determining $A_\mathrm{Fe}$ from each line. A microturbulence parameter $v_t$ was set in each case to remove trends between $\afe$(\fei) and line strength. These absolute abundances are consistent within the 1-$\sigma$ errors with excitation and ionization balance, as shown in Figure~\ref{f:reference}. The averages of the $A_\mathrm{Fe}$ values that we find for each star are also consistent, within the errors, with the $\feh$ values adopted from the literature, assuming $A_\mathrm{Fe}^\odot=7.45$, i.e., the meteoritic iron abundance \citep{lodders09}, or any modern determination of the solar iron abundance \cite[e.g., $\afe=7.50$ from][]{asplund09:review}. We stress that for our purposes the exact choice of solar abundances is inconsequential \citep{ramirez11,melendez12}.

\begin{table}
\caption{Adopted stellar parameters of the two reference stars HIP\,14954 (wm) and HIP\,74500 (mr).\tablefootmark{a}}
\centering
\scriptsize
\begin{tabular}{ccccccc}\hline\hline
Star       & $\teff$ & $\logg$ & $\feh$ & $v_t$ & \multicolumn{2}{c}{$\afe$} \\
           & K       & [cgs]   &        & \kms  & \fei   & \feii  \\ \hline
HIP\,14954 & 6150    & 4.15    & +0.20  & 1.4  & $7.64\pm0.06$ & $7.71\pm0.07$ \\
HIP\,74500 & 5750    & 4.30    & +0.25  & 1.1  & $7.70\pm0.08$ & $7.74\pm0.10$ \\ \hline
\end{tabular}
\label{t:reference}
\tablefoot{\small
\tablefoottext{a}{The $\teff$, $\logg$, and $\feh$ values listed here are fixed, therefore no error estimates are needed. The $v_t$ value corresponds to the microturbulence parameter that, within 0.1\,\kms, removes trends between line strength and iron abundance. The error bars for the iron abundances are the 1-$\sigma$ line-to-line scatter given those fixed stellar parameters.}
}
\end{table}

The absolute parameters adopted for the two reference stars are listed in Table~\ref{t:reference}. Absolute parameters for the rest of our sample stars, computed using these two references and the relative parameters derived as in Section~\ref{s:atmospheric_parameters}, are given in Table~\ref{t:pars} (columns 2--5). These values were employed hereafter to derive abundances for the other elements. We emphasize that these absolute parameters are only as good as those of the reference stars chosen. However, the relative values, which are listed in columns 6--11 of Table~\ref{t:pars} are much more reliable and have very high precision.

\begin{table*}
\caption{Atmospheric parameters\tablefootmark{a}}
\centering\tiny
\begin{tabular}{lccrcrcrcrc}\hline\hline
HIP & $\teff$ & $\logg$ & $\feh$ & $v_t$ & $\Delta\teff$ & error & $\Delta\logg$ & error & $\Delta\feh$ & error \\
 & (K) & [cgs] & & (\kms) & (K) & (K) & [cgs] & [cgs] & & \\ \hline
\input{pars_small.tex}
\hline
\end{tabular}
\tablefoot{\small
\tablefoottext{a}{The $\Delta$ values were measured using our spectra, and are given here relative to the reference stars HIP\,14954 (wm) and HIP\,74500 (mr). The $\teff$, $\logg$, $\feh$, and $v_t$ values were determined by adding these $\Delta$ values to the adopted parameters of the reference stars, as given in Table~\ref{t:reference}.}
}
\label{t:pars}
\end{table*}

The errors of the relative parameters listed in Table~\ref{t:pars} were determined as follows. The standard excitation/ionization balance procedure described before provides formal errors in the parameters derived. In our case they are measured by propagating the uncertainty in the slopes of the final iron abundance versus EP and line-strength relations as well as the 1-$\sigma$ line-to-line scatter of the mean \fei\ minus \feii\ iron abundances. For a given calculation (i.e., for one pair of spectra), these errors are larger than those given in Table~\ref{t:pars} because self-improvement reduces the errors. If the data were perfectly homogeneous, and no inter-dependencies existed, the errors of the self-improved parameters would be scaled down as $n^{-1/2}$, where $n$ is the number of spectra analyzed. In our case that would give errors in $\teff$ as small as a few degrees, which is not realistic. The relative errors computed within the self-improvement scheme, however, are accurate. Therefore, the errors listed in Table~\ref{t:pars} were obtained by multiplying the formal errors from the self-improved parameters by a scale factor that makes the average errors of the sample identical to those derived from the simulation discussed in Section~\ref{s:atmospheric_parameters}.

Abundances of 18 elements other than iron were measured using $EW$ analysis with MOOG. The linelist adopted is from the \cite{ramirez09} work, but the number of features is smaller due to the more limited wavelength coverage of the data employed here. Hyperfine structure was taken into account for \ion{V}{i}, \ion{Mn}{i}, \ion{Co}{i}, \ion{Cu}{i}, and \ion{Ba}{ii} using the wavelengths and relative $\log gf$ values from the Kurucz atomic line database.\footnote{http://kurucz.harvard.edu/linelists.html} Our linelist (including the iron lines) is given in Table~\ref{t:linelist}.

\begin{table}
\caption{Line list\tablefootmark{a}}
\centering\tiny
\begin{tabular}{ccccc}\hline\hline
Species & $T_\mathrm{C}$ & Wavelength & EP   & $\log gf$ \\ 
        & (K)            & (\AA)    & (eV) &           \\ \hline
\input{ll_small.tex}
\hline
\end{tabular}
\tablefoot{\small
\tablefoottext{a}{Condensation temperatures adopted are given with the first line of each species. The components employed for lines where hyperfine structure was taken into account are also listed (for example the 6039.7\,\AA\ V\,\textsc{i} line).}
}
\label{t:linelist}
\end{table}

Similar to the case of stellar parameter determination, the absolute abundances were used to construct a matrix of relative abundances for each element. This means that we determined differential abundances on a line-by-line basis of every star relative to all others and employed self-improvement to obtain unique, consistent differential values and to minimize the impact of outliers. Table~\ref{t:abundances} contains our final relative abundances. In order to transform those values into the more traditional [X/Fe] abundance ratios, those of the reference stars (on an absolute scale or relative to solar) must be first determined. That may be useful for Galactic chemical evolution (GCE) studies, but here we are only interested in relative element depletions, free from GCE effects since each sample span a very small [Fe/H] range.

\begin{table}
\caption{Relative abundances\tablefootmark{a}}
\centering\scriptsize
\begin{tabular}{lrcrcrcc}\hline\hline
\input{abundances_small.tex}
\hline
\end{tabular}
\tablefoot{\small
\tablefoottext{a}{The $\Delta$ values listed here are relative to the reference stars HIP\,14954 for the wm sample and HIP\,74500 for the mr sample. The $\sigma$ columns correspond to a simple line-by-line scatter and are not necessarily representative of the true error of our relative, strict differential abundance measurements, which we estimate to be around 0.01\,dex.}
}
\label{t:abundances}
\end{table}

\subsection{Mass and convective envelope size} \label{s:cz}

Using the atmospheric parameters listed in Table~\ref{t:pars}, we estimated the stars' masses employing the Yonsei-Yale grid of theoretical isochrones \citep{yi01,kim02}. The details of our procedure to determine these masses have been described multiple times \cite[see, for example, Section 3.2 in][but also \citealt{chaname12,melendez12}]{ramirez13:thin-thick}, and will not be repeated here. In summary, a mass probability distribution is calculated by comparing the location of isochrone points in the $\teff$, $\logg$, $\feh$ space to the parameters measured in each star. The peak of that distribution is adopted as the most likely mass of the star. We find that the average mass of the wm sample stars is 1.2\,$M_\odot$, while that for the mr stars is 1.1\,$M_\odot$. The dotted lines in Figure~\ref{f:sample} illustrate that these masses are indeed representative of their respective samples.

The Yonsei-Yale isochrones were computed using solar-scaled compositions, adopting the solar mixture given by \cite{grevesse93} and a mixing length parameter consistent with these abundances at solar age. Our model atmosphere analysis uses the Kurucz ``odfnew'' grid, which adopts solar abundances from \cite{grevesse98}. This inconsistency in the adopted solar abundances could be one of the reasons why the masses and $\logg$ values inferred using the stars' measured {\it Hipparcos} parallaxes are slightly offset relative to those determined using only our high resolution spectra. We find that the masses ($\logg$ values) obtained using parallaxes are $0.02\pm0.02$ ($0.04\pm0.04$) higher (lower) than those estimated from spectroscopy alone. Note that these systematic uncertainties are comparable in size to our formal errors. Nevertheless, it is also important to point out that stellar evolution calculations are highly model dependent, and that these differences in inferred stellar parameters could also be due to the many other factors involved in the modeling of both stellar atmospheres and internal evolution.

Using the pre-main-sequence (PMS) tracks of \cite{siess00}, we determined the typical size of the convective envelopes in our stars. We obtained them by interpolation in mass and metallicity at the zero-age-main-sequence. The metallicity adopted for the typical wm star is $\feh=0.1$ while that for the typical mr star is $\feh=0.2$ (see Figure~\ref{f:sample}). Considering the average masses and metallicities quoted above, we find that for the wm and mr samples the typical convective envelope masses are 0.005 and 0.017\,M$_\odot$, respectively, while for the Sun it corresponds to 0.023\,M$_\odot$. Note that the convective envelope mass of the mr sample is smaller than the solar one. This is due to the fact that a typical star from this sample is somewhat more massive than the Sun, and the decrease due to the larger mass is more important than the small increase due to the higher metallicity.

\subsection{Age and chromospheric activity}

The exact same procedure describe above to estimate the stars' masses was employed to calculate their ages. Although our sample stars are on the main-sequence, which typically prevents a precise measurement of isochrone ages, the extreme high-precision of our derived atmospheric parameters allows us to derive reasonably reliable relative ages. Of course, our absolute ages are still highly model-dependent, and the use of a different isochrone set will result in different ages. Nevertheless, we will use the age information to separate young stars from old stars and to sort them according to evolutionary state. For these purposes, our precise relative results are sufficient.

In addition, to investigate the potential effects of stellar activity on our results, we computed the chromospheric activity index $\log R'_\mathrm{HK}$ as follows. First, the fluxes in the cores of the \ion{Ca}{II} H and K lines at 3934 and 3968~\AA\ were measured using triangular passbands 1\,\AA\ wide. Pseudo-continuum fluxes were measured using the $3925\pm5$ and $3980\pm5$~\AA\ windows. The ratio of the \ion{Ca}{II} H and K fluxes to the pseudo-continuum fluxes provides us with an instrumental $S_\mathrm{inst}$ value for each star. To standardize these measurements and place them into the Mount Wilson system, we searched for previously published $S_\mathrm{MW}$ values for our sample stars in the catalogs by \cite{duncan91,henry96,wright04,gray06,jenkins06,jenkins11} and \cite{cincunegui07}. A linear fit of $S_\mathrm{inst}$ versus $S_\mathrm{MW}$ for the stars with previously published $S_\mathrm{MW}$ measurements allowed us to transform all our $S_\mathrm{inst}$ measurements into $S_\mathrm{MW}$ values. The fits were made independently for the wm and mr samples. Combining all stars resulted in a less precise fit. $B-V$ colors listed in the Hipparcos catalog were then employed to transform $S_\mathrm{MW}$ into $\log R'_\mathrm{HK}$ using equations 9 to 12 in \cite{wright04}. Our $\log R'_\mathrm{HK}$ measurements show good agreement with previously published values (as given in the references cited above). For the wm sample the mean difference is $-0.009\pm0.048$ while that for the mr sample is $0.005\pm0.043$. Thus, our $\log R'_\mathrm{HK}$ values have errors of order 0.04--0.05.

Our derived masses, ages, and the chromospheric activity index $\log R'_\mathrm{HK}$ are listed in Table~\ref{t:magerhk}. The $\pm2\sigma$ values represent 95\,\% confidence intervals.

\begin{table}
\caption{Mass, age, and chromospheric activity index}
\centering\scriptsize
\begin{tabular}{cccccccc}\hline\hline
HIP & Mass & $-2\sigma$ & $+2\sigma$ & Age & $-2\sigma$ & $+2\sigma$ & $\log R'_\mathrm{HK}$ \\ 
&$(M_\odot)$&$(M_\odot)$&$(M_\odot)$&(Gyr)&(Gyr)&(Gyr)&\\ \hline
\input{agerhk_small.tex}\hline
\end{tabular}
\label{t:magerhk}
\end{table}

\section{Depletion patterns} \label{s:discussion}

\subsection{Pristine versus depleted stars}

In M09's solar twin experiment, the reference star for chemical abundances was the Sun. Refractory element depletions could be attributed to planet formation because we know that the Sun hosts a planetary system that includes rocky objects. In R11's 16\,Cygni work, one of the stars in the binary system is known to host a gas giant planet whereas the other one does not show evidence of sub-stellar mass companions. In both cases it was straightforward, based on previous knowledge, to determine which star is expected to show element depletions. In our case, where large samples of stars are analyzed and no previous knowledge of their complete planet properties is available, the solution to the problem is not that simple. Although we have some information on which stars host planets, we do not know for sure whether some, most, or indeed all of our targets host smaller planets. An indirect approach, with some underlying reasonable assumptions, should therefore be employed.

We are mainly interested in refractory element depletions, suggested to be signatures of rocky planet formation (as in M09). The signature of gas giant formation suggested by R11 is not possible to detect in isolated stars because both refractory and volatile elements are expected to be depleted. In other words, the latter results in a constant metallicity offset (i.e., all elements decrease by the same amount) and it is therefore not trivial to disentangle GCE effects from the postulated effect of gas giant planet formation.

\begin{figure}
\includegraphics[bb=65 365 450 1140,width=9.4cm]
{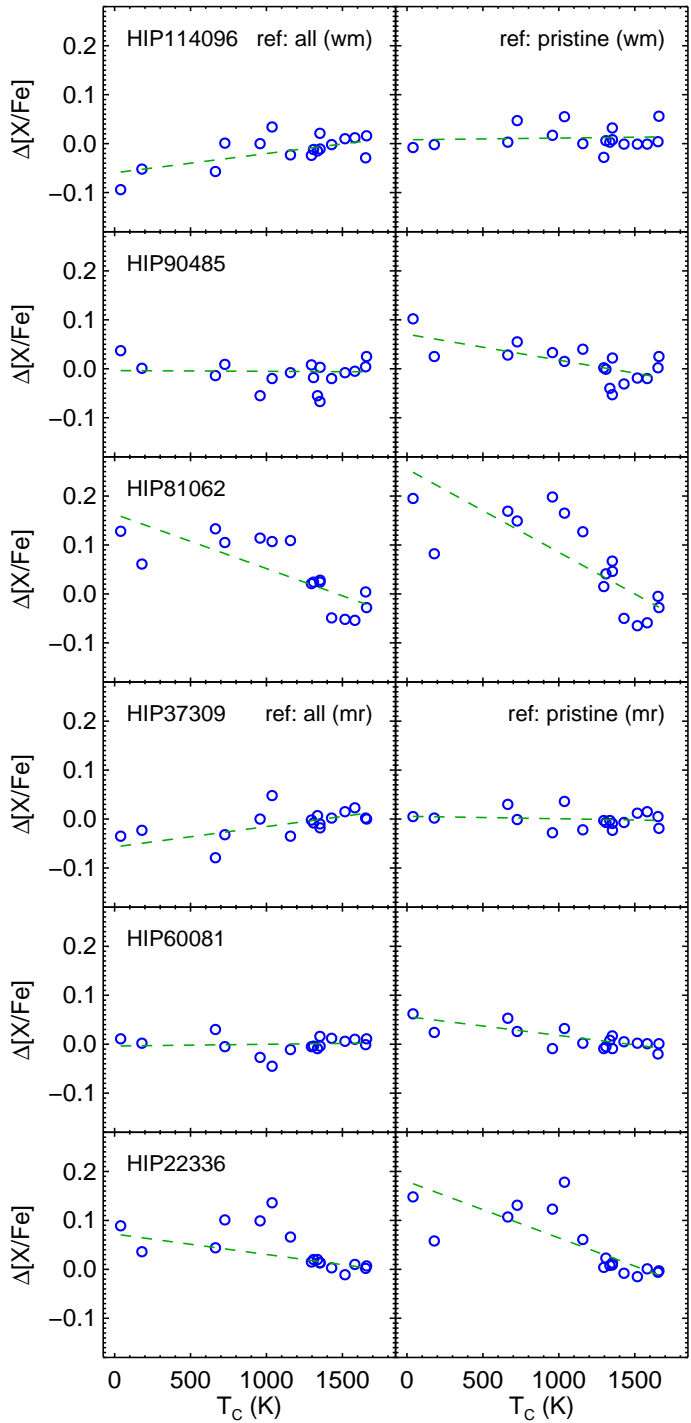}
\caption{Left panels: Relative [X/Fe] abundance ratios of a few representative sample stars with respect to the average values of all other stars as a function of the elements' condensation temperature ($\tc$). Right panels: As in the left panels, but with respect to the average values of stars that do not exhibit refractory element depletion (i.e., stars with ``pristine'' composition). Dashed lines are linear fits to the data. The top (bottom) three panels correspond to stars in our warm (metal-rich) sample.}
\label{f:xfe_tc}
\end{figure}

The problem is how to define a sample of stars with ``pristine'' composition, i.e., non-refractory-element-depleted stars. To achieve this goal, we inspected $\Delta\xfe$ versus $\tc$ plots, where $\Delta\xfe$ is the average relative abundance ratio of element X to Fe of a given star with respect to all others,\footnote{In this context we imply: ``relative to the {\it average} abundance ratios of all other stars in that sample.''} and $\tc$ is the 50\,\% condensation temperature of element X. We employed the $\tc$ values computed by \cite{lodders03} for a solar composition gas (these values are listed in Table~\ref{t:linelist}). Refractory (volatile) elements have high (low) $\tc$. A few representative examples of the $\Delta\xfe$ versus $\tc$ relations are shown in Figure~\ref{f:xfe_tc}, left panel.

The stars shown in Figure~\ref{f:xfe_tc} illustrate the general behavior of the sample with regards to refractory element depletions. On the left panels we show the average abundance ratios of a given star relative to all others in that sample. Three cases are plotted for each sample. The first ones (HIP\,114096 for the wm sample and HIP\,37309 for the mr sample) are stars that relative to all others present a positive $\Delta\xfe$ versus $\tc$ slope. One way of interpreting this observation is that these objects are the least depleted in refractories, since they are overabundant in those elements.\footnote{Although $\Delta\xfe=0$ for the most refractory elements in these cases, we note that this is because Fe is being used as reference in the abundance ratio [X/Fe], and Fe is a refractory element, with $\tc=1334$\,K. Had we chosen a volatile as reference, these abundance ratios would be positive. We continue to use Fe as reference because it has the largest number of features available in our spectra, and we are therefore able to measure, internally, very precise Fe abundances.} The stars that appear next in Figure~\ref{f:xfe_tc} have near zero slope (wm: HIP\,90485 and mr: HIP\,60081). These are average stars with regards to the amount of refractory element depletion. Finally, HIP\,81062 (wm) and HIP\,22336 (mr) are stars with very negative slopes, which implies that they are the most refractory-element-depleted stars from their respective samples.\footnote{It should be noted that for some stars the element-to-element scatter in Figure~\ref{f:xfe_tc} appears higher than our error estimate of 0.02\,dex. This is not evidence that our errors are underestimated, because the planet signature hypothesis does allow for an intrinsic element-to-element scatter. Also, the abundances of some of the elements included in this work, namely Zn and Ba, are known to exhibit large star-to-star scatter at constant $\feh$ \cite[see, e.g., Figure~1 in][]{ramirez09}.}

\begin{figure}
\includegraphics[bb=80 370 420 715,width=9.3cm]
{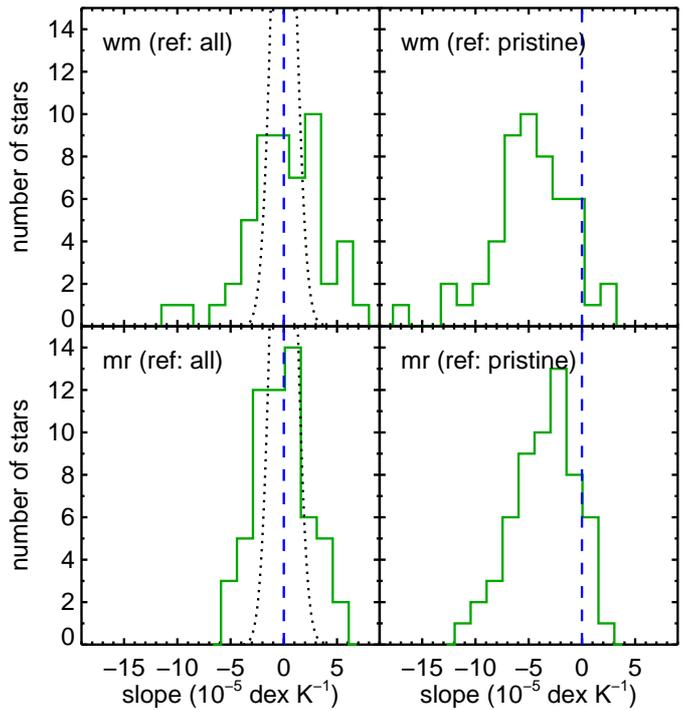}
\caption{Left panels: distribution of $\Delta\xfe$ versus $\tc$ slopes when the abundance ratios are measured with respect to the average of all stars. The dotted lines represent the distribution of slopes of data with no depletions and pure observational noise. Right panels: as in the left panels, but for the case when the abundance ratios are measured with respect to the average of stars with pristine composition. The dashed line is at zero slope. The top (bottom) panels correspond to our warm (metal-rich) sample.}
\label{f:slopes}
\end{figure}

The distribution of slopes of the $\Delta\xfe$ versus $\tc$ relations naturally center around zero because the reference for $\Delta\xfe$ is the average of all stars (see Figure~\ref{f:slopes}, left panels). Stars with a positive $\Delta\xfe$ versus $\tc$ slope are more refractory rich than the average star of the sample while those with negative slopes can be interpreted as refractory-element-depleted stars.

It is important to check that the slope distributions shown in Figure~\ref{f:slopes} are not simply due to errors in the relative abundance determinations. A somewhat conservative estimate for the latter is 0.02\,dex (we expect most abundance ratios to be precise at the 0.01\,dex level). We generated 10,000 $\Delta\xfe$ versus $\tc$ relations, with the $\Delta\xfe$ values taken from a Gaussian distribution of 0.02\,dex of standard deviation and zero mean. We computed the $\Delta\xfe$ versus $\tc$ slopes for each of these relations and determined their distributions, normalizing them to have an area equal to the number of stars in each of our real samples. These distributions, which are shown with dotted lines in Figure~\ref{f:slopes}, left panels, are clearly too narrow compared to those measured using the real data. This ensures that our analysis is based on measurements of actual element depletions and not those of random observational noise.

In the context of the M09 hypothesis, stars with the most positive slopes can be thought of as stars with pristine composition, i.e., objects that have not been depleted in refractory elements by the process of terrestrial planet formation. These stars can be considered better references in our analysis because we are interested in finding refractory-element-depleted stars, not with respect to the average star in the sample, but with respect to stars which have a chemical composition representative of the gas cloud which formed them, and was not affected by the process of planet formation. Thus, for each of our samples, we define a sub-sample of pristine stars which consists of the 15\,\% of stars with the most positive $\Delta\xfe$ versus $\tc$ slopes.

In a next step, we re-computed $\Delta\xfe$ values, but this time using the pristine sample as reference (as opposed to all other stars), and inspected the resulting $\Delta\xfe$ versus $\tc$ relations (Figure~\ref{f:xfe_tc}, right panels). As expected, the distribution of $\Delta\xfe$ versus $\tc$ slopes is now skewed towards negative values (see Figure~\ref{f:slopes}, right panels), since, by definition, all stars that do not have pristine composition have been refractory-element-depleted by a certain amount. Note that this is not simply a shifted distribution relative to those shown on the left panels of Figure~\ref{f:slopes}, because the corresponding $\Delta\xfe$ average abundance ratios use a different set of references (all other stars or only those with pristine composition). The slope distribution of the metal-rich solar-analog sample appears to have a longer negative tail when using the pristine sample as reference, but this is not statistically significant (this is made clear if one compares the extent of this tail with the amplitude of the pure noise distribution shown in the left panels of Figure~\ref{f:slopes}).

Hereafter only the pristine samples are used as reference to compute the relative abundance ratios $\Delta\xfe$ and $\Delta\xfe$ versus $\tc$ slopes.

\subsection{Stars known to host planets}

\begin{figure}
\includegraphics[bb=90 370 420 715,width=9.2cm]
{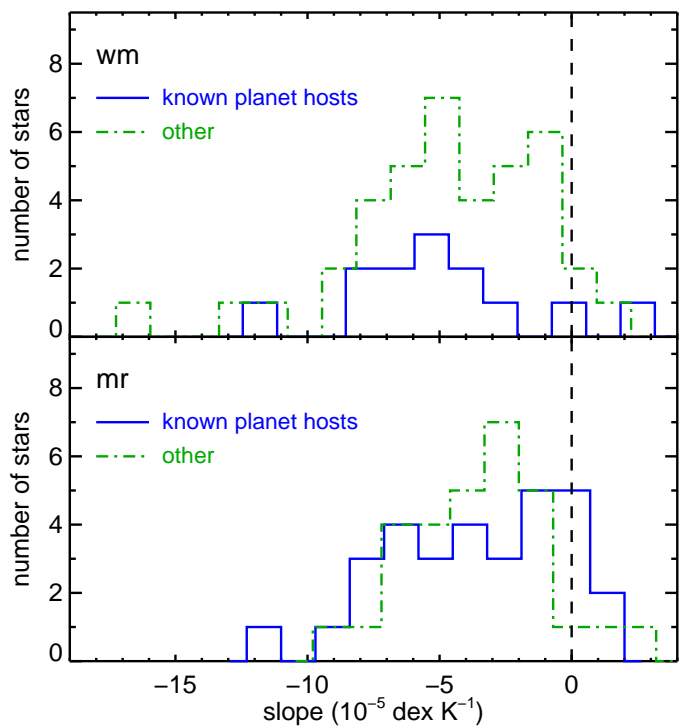}
\caption{The distribution of $\Delta\xfe$ versus $\tc$ slopes for stars with known planets (solid line histogram) is compared to that of the rest of stars in each sample (dot-dashed line histogram). The dashed line is at zero slope. The top (bottom) panel corresponds to our warm (metal-rich) sample.}
\label{f:slope_ph}
\end{figure}

According to the M09 hypothesis, those stars with the most negative $\Delta\xfe$ versus $\tc$ slopes should have formed rocky material with the largest total mass (within their respective samples). In Figure~\ref{f:slope_ph} we compare the distribution of $\Delta\xfe$ versus $\tc$ slopes for known planet-hosts in our sample to that of the rest of stars. We do not observe an offset in the distributions, particularly with the planet-host sample shifted towards more negative slopes, as one would expect if they host terrestrial planets. However, since most of the planets known around our sample stars are gas giant or Neptune-size objects, it is not necessarily the case that they will present refractory-element depletions like the Sun. Although for simplicity we use the term refractory-element depletion, we should keep in mind that in fact the observation suggests a deficiency of refractories relative to volatiles. Thus, even if these large planets have massive rocky cores, their volatile content may be also high, which would flatten the $\Delta\xfe$ versus $\tc$ relations, as indeed seems to be the case for the gas giant planet orbiting 16\,Cygni\,B (R11).

\begin{figure}
\includegraphics[bb=75 370 420 715,width=9.3cm]
{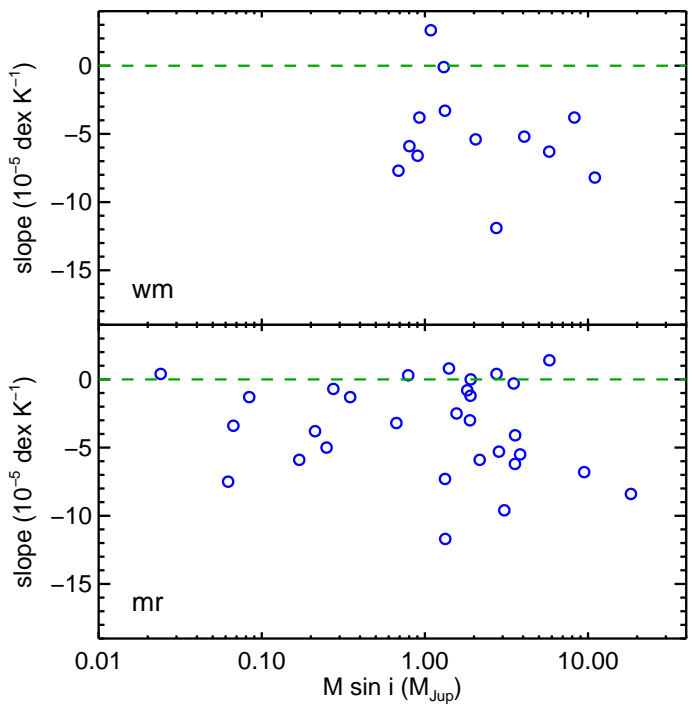}
\caption{$\Delta\xfe$ versus $\tc$ slopes, where the abundance ratios are measured relative to the average of stars with pristine composition, as a function of minimum mass of the planet hosted (for multi-planet systems, we consider only the most massive planet). The dashed line is at zero slope. The top (bottom) panel corresponds to our warm (metal-rich) sample.}
\label{f:slope_ph_mass}
\end{figure}

Another way of analyzing these results is by looking at a scatter plot of the $\Delta\xfe$ versus $\tc$ slopes versus planet mass. In a simplistic interpretation of the M09 hypothesis, one would expect more negative slopes for smaller planets. Figure~\ref{f:slope_ph_mass} shows no observational evidence for the previous statement. There is, in fact, no obvious correlation between planet mass and slope value for either one of our samples. However, we should emphasize that we are typically dealing with big planets, with the lower envelope in planetary mass being about the mass of Neptune for the metal-rich sample and about 1 Jupiter mass for the warm sample. In both cases the range in planet mass extends to about 10 Jupiter masses. Thus, these planets are not necessarily expected to present a clear specific signature with condensation temperature. This is in line with  what was found by R11 in the analysis of the 16 Cygni system, where the planet-host star only showed an overall depletion in all chemical elements, without any specific trend with condensation temperature.

Other scenarios could explain the lack of correlation between slope value and presence of gas giant planets. Very early accretion of gas deficient in planet material, for example, will not be able to imprint a chemical signature on the star's photosphere due to the large mass of its convective envelope. Also, late accretion of rocky material can potentially erase any signatures imprinted early on.

The least massive planets in our sample orbit stars in our metal-rich group, as can be seen in Figure~\ref{f:slope_ph_mass}. Probably only one of them is a super-Earth \cite[the 7.4\,$M_\oplus$ planet orbiting HIP\,1499,][]{rivera10} and could therefore be expected to reveal the signature of rocky element depletion suggested by M09. Interestingly, this object belongs to our pristine composition sub-sample of metal-rich stars, with a $\Delta\xfe$ versus $\tc$ slope close to zero, in apparent stark contradiction with the M09 hypothesis. The four planets with minimum mass below $0.1\,M_\mathrm{Jup}$, some of which may have a significant mass of rock, do not appear to prefer a certain $\Delta\xfe$ versus $\tc$ slope value, high or low, but roughly span the range covered by the entire sample of planet-hosts. We should note, however, that the term super-Earth does not mean that the planet has a rocky composition, only that it is significantly more massive than the Earth, so a comparison with the signature imprinted by rocky planets could be unfair. At this point it is important to remember that we do not know the exact bulk chemical composition of the Solar System gas giants and that the situation is even worse regarding the super-Earths. Thus, we cannot claim for sure that super-Earths have to present the chemical signature of rocky planets suggested by M09.

\subsection{Amplitude of the depletions}

The planet signatures suggested by M09 and R11 are imprinted on the stars' convective envelopes. Thus, their amplitudes should be sensitive to the total mass of convective envelopes in the stars analyzed. In particular, in main-sequence stars warmer than the Sun, which have thinner, less massive convective envelopes, the dilution of the chemical planet signature is expected to be less important than in the Sun. If everything else is the same, this would result in a larger amplitude of refractory-element depletion because we measure surface composition and the photospheric material is expected to be well mixed with the star's convective envelope gas.

To determine the amplitude of refractory-element depletion, we selected in each sample the 15\,\% of stars with the most negative $\Delta\xfe$ versus $\tc$ slopes and computed their average $\Delta\xfe$. This selection gives us confidence that if the refractory-element depletion is due to planet formation (as hypothesized in the solar case), we are comparing the most reliable rocky planet host candidates to the stars that most likely did not form any of those objects. Note that in this procedure, we are not really excluding stars from the calculations, because the pristine sample was defined using all stars. Thus, this is not a biased comparison that uses a selected group of stars from our complete samples, but our best attempt at reducing the uncertainties.

\begin{figure}
\includegraphics[bb=70 370 420 715,width=9.3cm]
{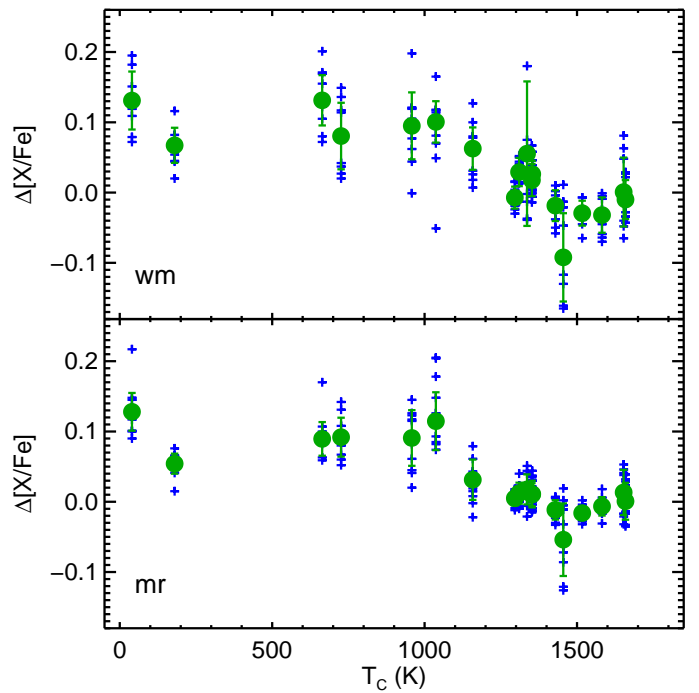}
\caption{Crosses are the relative abundance ratios of the most refractory-element-depleted stars in our samples, using the pristine stars as reference for the $\Delta\xfe$ values. Filled circles with error bars represent the weighted mean and standard deviation of the data plotted with crosses. Thus, they illustrate the amplitude of refractory-element depletions. The top (bottom) panel corresponds to the warm (metal-rich) sample.}
\label{f:depletions}
\end{figure}

Figure~\ref{f:depletions} shows the $\Delta\xfe$ versus $\tc$ relations for the most refractory-element-depleted stars in each sample. The star-by-star data are shown with crosses, and the element-by-element averages of all these stars are plotted with filled circles (the error bars correspond to the 1-$\sigma$ star-to-star scatter). The maximum amount of refractory-element depletion is about 0.15\,dex, independently of the sample. If this is due to a depletion of elements in the stars' convective envelopes, it is difficult to reconcile the latter observation with the fact that stars in our warm sample have present-day convective envelopes that are about half as massive as those of our metal-rich sample stars. This, of course, assumes that the amount of rocky material formed around the most refractory-element-depleted stars is the same in both types of stars. If, on the other hand, stars in our metal-rich sample are able to form more terrestrial planets and meteorites than those in our warm sample, the dilution effect could be compensated.

\subsection{Comparison to M09's solar twin data} \label{s:m09data}

\begin{figure}
\includegraphics[bb=90 370 450 985,width=9.3cm]
{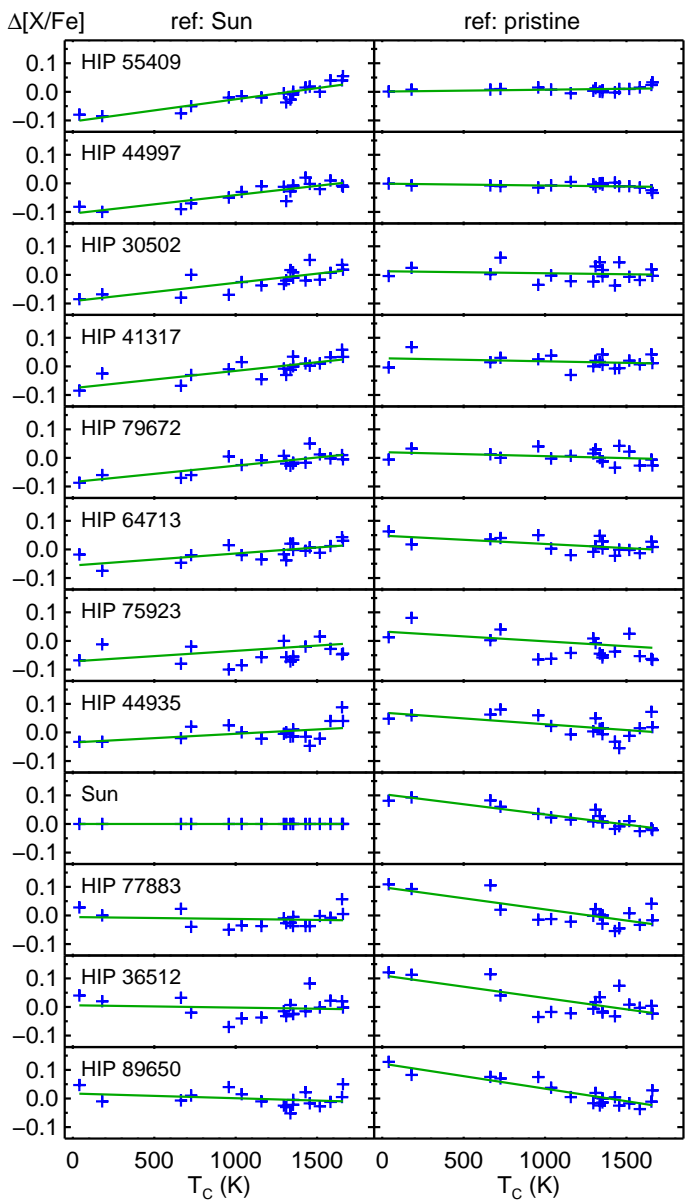}
\caption{Left panels: Crosses are the [X/Fe] versus $\tc$ relations for 11 solar twin stars according to Mel\'endez et~al.\ (2009). Solid lines are linear fits to the data. Right panels: as in the right panel, but for $\Delta\xfe$, where the chemical abundance reference is not the Sun, but the average of the two least refractory-element-depleted (``pristine'') solar twins.}
\label{f:m09}
\end{figure}

In order to place our results in the context of M09's solar twins work, we must first re-assess their data. M09 determined [X/Fe] abundance ratios using the solar spectrum as reference. These values are plotted as a function of $\tc$ in Figure~\ref{f:m09}, left column.\footnote{To remain consistent with the rest of our calculations, we only take into account the elements employed in this paper. M09 used more chemical elements than we had available for this work.} The stars have been sorted so that those with the most positive $\xfe$ versus $\tc$ slope are shown in the top panels. Note that the Sun is neither the most refractory-element depleted star in this sample nor the star with the most pristine composition. Since the M09 work includes data for 12 stars (11 solar twins and the Sun), picking the two with the most positive and most negative $\xfe$ versus $\tc$ slopes is equivalent to determining the $\sim15$\,\% most pristine and most refractory-element depleted stars, respectively.

\begin{figure}
\includegraphics[bb=82 370 420 573,width=9.3cm]
{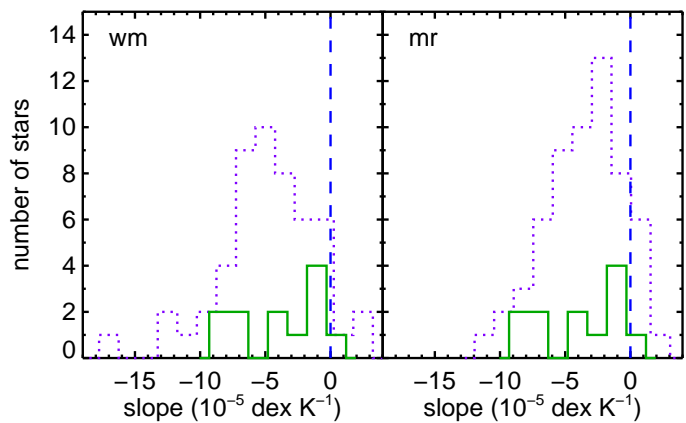}
\caption{The distribution of $\Delta\xfe$ versus $\tc$ slopes of solar twin stars in the M09 work (solid line histogram) is compared to that of our warm (left panel) and metal-rich (right panel) samples. The dashed line is at zero slope.}
\label{f:slopes_m09}
\end{figure}

The right panels of Figure~\ref{f:m09} show the $\Delta\xfe$ versus $\tc$ relations, where the reference for chemical abundances is now the average of the two solar twin stars with the most pristine composition (HIP\,55409 and HIP\,44997). The nature of these data is equivalent to that of our Figure~\ref{f:xfe_tc}, right panels. The solid lines over-plotted are linear fits to the $\Delta\xfe$ versus $\tc$ data (for simplicity, all data points were given equal weight in the fitting procedure; the few elements with large abundance uncertainties in the M09 work were already removed because they are not included in the present work). The distribution of the slopes of these fits is compared to that of our warm and metal-rich samples in Figure~\ref{f:slopes_m09}.

\begin{figure}
\includegraphics[bb=65 370 390 573,width=9.3cm]
{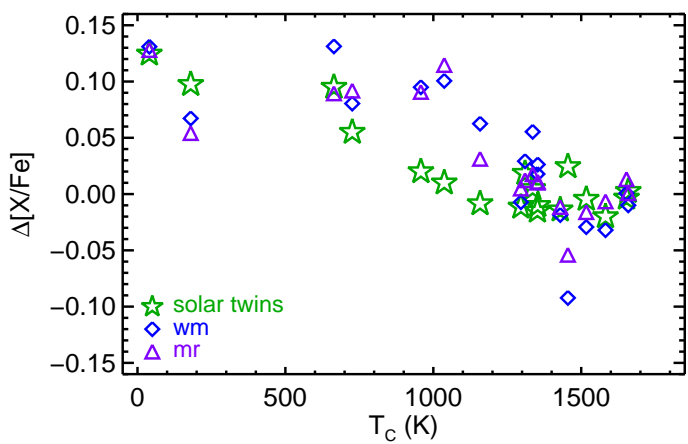}
\caption{Average difference in [X/Fe] between the most and least refractory-element-depleted stars in three samples as a function of $\tc$: M09's solar twins (five-pointed stars), and our warm (diamonds) and metal-rich (triangles) stars.}
\label{f:m09depletion}
\end{figure}

Although, based on the discussion above regarding the distribution of $\Delta\xfe$ versus $\tc$ slopes, one may expect the maximum amplitude of refractory-element depletion to be larger in our warm and metal-rich samples (because the histograms have a larger span and these amplitudes are determined by comparing the most refractory-element depleted to the most pristine stars), in fact there is not a noticeable difference between the three samples studied. Figure~\ref{f:m09depletion} shows the amplitude of refractory-element depletions in our warm (diamonds) and metal-rich (triangles) samples (these are the same data plotted in Figure~\ref{f:depletions} with filled circles). The corresponding data for solar twins are shown with five-pointed stars. The $\Delta\xfe$ values for solar twins in Figure~\ref{f:m09depletion} correspond to the difference in abundance ratios between the two most and two least refractory-element depleted stars in that sample (i.e., the $\sim15$\,\% of stars in the extremes of the slope distributions, exactly as adopted in our work).

Figure~\ref{f:m09depletion} shows that refractory-elements are depleted by about 0.15\,dex with respect to volatiles regardless of the sample analyzed, solar twins included. However, there is a notable difference between the solar twin data and that of our warm and metal-rich samples at $\tc\sim1000$\,K, where elements Na ($\tc=958$\,K), Cu ($\tc=1037$\,K), and Mn ($\tc=1158$\,K) are found. For these elements, we find high $\Delta\xfe$ values while M09's data suggest values near zero. Thus, even though the amplitude of refractory-element depletions is very similar, the morphology of the $\Delta\xfe$ versus $\tc$ relation is not the same.

\begin{figure}
\centering
\includegraphics[bb=65 375 390 855,width=9.3cm]
{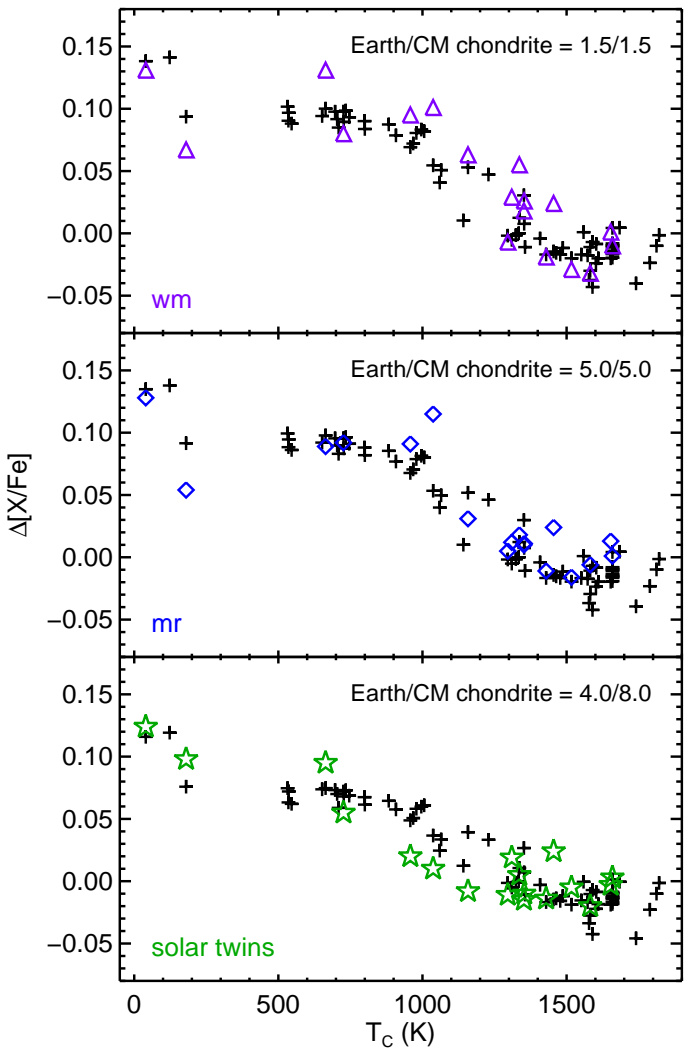}
\caption{The triangles, diamonds, and stars show the depletion patterns of Figure~\ref{f:m09depletion}, while the crosses represent the effect of a mixture of Earth-like and meteoritic-like material on the convective envelopes of the different samples. The mass ratio of Earth-like and meteoritic-like material is given on the top right corner of each panel; the values given correspond to Earth masses ($M_\oplus$). The convective envelope masses adopted are the present-day ones of a typical star in each sample.}
\label{f:rocks}
\end{figure}

In order to model the depletion of refractory elements in the warm, metal-rich, and solar twin samples, we followed the same approach as \cite{chambers10}. In summary, we calculated the change in chemical composition that a star experiences due to the formation of rocky material for a given mixture of Earth-like and meteoritic-like material, finding the best combination by comparing the results of this calculation to the observations. The best fits to the data for this experiment are shown in Figure~\ref{f:rocks}. Typically, from a few (for the wm sample) to $\sim 10\,M_\oplus$ (solar twins) of refractory-rich rocky material would need to be removed from the stellar convection zones to produce the observed abundance pattern; we note that these estimates differ somewhat from those in M09 since here we are trying to reproduce the signatures of the most depleted stars rather than the average of all solar twins as in M09. There is a correlation between the size of the convective envelope and the amount of rocky material needed to explain the observed abundance pattern, with the F stars requiring less amounts and the solar twins the most. It is interesting to note that although both warm and metal-rich stars seem to require an equal mixture of terrestrial-like and meteoritic-like material (but more quantity for the metal-rich stars), the solar twins need about twice as much meteoritic-like material than Earth-like material. This is a direct consequence of the difference in morphology of the $\Delta$[X/Fe] versus $\tc$ relations discussed before (cf.\ Figure~\ref{f:m09depletion}).

The calculations described above employed the convective envelope size of present-day stars. However, we know that Sun-like stars are born fully convective and that their convective envelopes become thin as planets form. Thus, an alternative explanation for Figure~\ref{f:rocks} could be that planets form on shorter time-scales around the warm stars (when the convective envelope is massive), somewhat longer time-scales for metal-rich stars, and on very long time-scales for solar twins (essentially when the star's convective envelope has reached its final size). Interestingly, observations suggest that disk lifetimes are shorter around higher mass objects, likely owing to faster accretion and more intense radiation \citep{williams11}. Since the disk lifetime sets a limit on the time available for planet formation, the shorter disk lifetime around the more massive F dwarfs is in line with the lower mass of refractory-elements needed to explain our observations.

\subsection{Dependence on stellar parameters}

\begin{figure*}
\centering
\includegraphics[bb=65 370 790 685,width=18cm]
{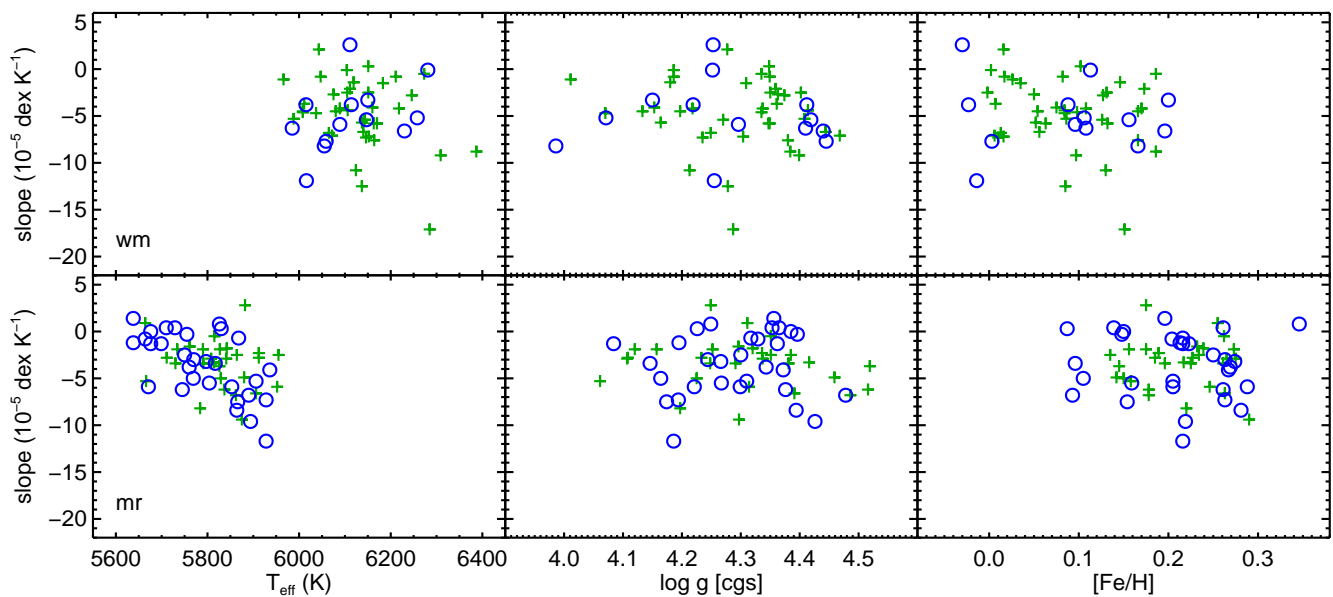}
\caption{$\Delta\xfe$ versus $\tc$ slopes as a function of atmospheric parameters for our warm (top panels) and metal-rich (bottom panels) samples. Known planet hosts are shown with open circles.}
\label{f:slopevspars}
\end{figure*}

Even though our samples cover small regions of stellar parameter space, which allows us to determine high-precision relative chemical abundances, we searched for correlations between the $\Delta\xfe$ versus $\tc$ slopes and $\teff$, $\logg$, and $\feh$, as shown in Figure~\ref{f:slopevspars}.

We do not detect any important correlations with either $\logg$ or $\feh$, but there is one between the slope values and $\teff$ for our metal-rich sample (see bottom left panel in Figure~\ref{f:slopevspars}). More objects with low refractory-element depletion or even pristine composition are found at lower $\teff$ ($\sim5600$\,K) in this group of stars. On the other hand, the most refractory-element depleted stars in our metal-rich sample have $\teff\sim5900$\,K. By comparing the slope versus $\teff$ relations of stars known to host planets (open circles) and the rest of objects (crosses), it appears that the former have a slight preference for more negative slopes. Note, however, that there are a few known planet-host stars with near-zero slope values (and in particular higher than most of the rest of stars) even at relatively high $\teff$ (e.g., at $\teff\sim5850$\,K). Thus, we should be careful and consider this small offset between known planet hosts and the rest of stars barely significant.

The $\Delta\xfe$ versus $\tc$ slope that we derive for the Sun is $-7.1\times10^{-5}$\,dex\,K$^{-1}$. If we include the Sun in Figure~\ref{f:slopevspars} it would appear close to the lower envelope of the slope versus $\teff$ scatter plot. In other words, for its $\teff$, the Sun would be one of the most refractory-element-depleted stars. However, this comparison does not take into account the fact that this sample of stars is significantly more metal-rich than the Sun so there may be a bias.

The correlation between $\Delta\xfe$ versus $\tc$ slope and $\teff$ is weak, or non-existent, for our warm sample. The direction of this weak correlation, however, is consistent with that observed in the metal-rich sample, i.e., more negative values with higher $\teff$. For $\teff<6200$\,K, planet-host stars seem to have slightly more negative slope values, but note that the star with the most pristine composition (i.e., the one with the highest slope value) is also a known planet host.

The fact that the maximum amplitude of refractory-element depletion is about the same for the three samples examined in this work (cf.\ Figure~\ref{f:m09depletion}) can be attributed to the observation that for the metal-rich sample, the stars with the largest depletion of refractories are also the warmest in their group, with $\teff$ values comparable to those of the coolest stars in our warm sample. The most refractory-element-depleted stars in the latter group span a range of $\teff$ values, all warmer than any star in the metal-rich group, but the maximum amplitude of refractory-element depletion in these stars is independent of $\teff$. On the other hand, as mentioned before, for its $\teff$ the Sun has a very low $\Delta\xfe$ versus $\tc$ slope (but still within the limits of the solar twin sample).


\subsection{The HARPS-GTO chemical analysis}

For many years, the HARPS-GTO high-precision planet-search program \cite[e.g.,][]{mayor03} has been monitoring a large sample of solar-type stars, leading to the discovery of an important number of extrasolar planets of various masses, including Neptune- and super-Earth-like planets. Equally important, in particular for the purposes of the investigation presented in this paper, is their determination of absence of sub-stellar mass companions to these stars (more accurately, the limits set on which types of planets cannot be present around those objects). The large number of spectra collected for each of the stars in this program and the long-term stability of the HARPS instrument has allowed them to perform high-precision chemical analysis for a variety of purposes \cite[e.g.,][]{santos04,ecuvillon06,israelian09,neves09,adibekyan13}. The works by \citet[hereafter GH13]{gonzalez-hernandez10,gonzalez-hernandez13} are particularly relevant here because they tackle the same problem addressed in this paper, but using the HARPS-GTO data.

Compared to the Sun, GH13 find that ``hot'' solar analogs with planets are slightly enhanced in refractories, and that the same can be said of their sample of ``single'' stars, but the enhancement is smaller for the latter. This apparently contradicts our hypothesis of refractory-element depletions being due to the formation of planets, but, similar to what we find in our work. we must keep in mind the fact that most of the planets found around the GH13 stars are gas giants and Neptune-like planets, whereas M09's hypothesis is related to terrestrial planets.

The $\feh$ coverage of the GH13 sample is wider than ours (0.8\,dex instead of our 0.2\,dex), which could result in larger systematic errors, particularly considering that their analysis is done differentially with respect to the Sun. The impact of Galactic chemical evolution (GCE) was taken into account in the GH13 work by fitting straight lines to the [X/Fe] versus [Fe/H] relations and removing those mean trends from the original abundance data. We avoided this approach because it could be that GCE effects are of similar amplitude compared to the element depletions due to planet formation. In other words, it may not be possible to trace GCE from [X/Fe] versus [Fe/H] plots because both [X/Fe] and [Fe/H] are affected by planet formation processes. Removing GCE effects in that manner may be resulting in a removal of the planet signature. Nevertheless, GH13 find similar results when restricting their sample to the narrow [Fe/H] range from +0.04 to +0.19, i.e., stars with and without detected planets in that narrow $\feh$ window still seem to be both slightly enhanced in refractories relative to the Sun.

By examining the $\tc$ abundance trends as a function of planet mass, GH13 find that stars hosting Neptune-like and super-Earths are in fact more depleted in refractory elements than stars hosting gas giants. However, GH13 point out that the statistics of low-mass planets is not ideal; only 4 Neptune-like planets and 8 super-Earths are found in their sample, but 21 gas giant planet-hosts are included. Thus, this agreement with the M09 hypothesis should be considered tentative at this point. This statement is further supported by their analysis of 10 stars that host super-Earths (two additional super-Earth planet hosts were added from their previous work on solar-analogs). The $\Delta\xfe$ versus $\tc$ plots for these stars do not reveal any consistent pattern. GH13 find that there is a roughly equal number of super-Earth planet-hosts exhibiting refractory-element depletions and enhancements (relative to the Sun). Also, they find that pairs of stars that host super-Earths of similar minimum mass can have widely different amounts of refractory-element depletion.

As acknowledged by GH13, although the latter appears to contradict M09's hypothesis, we should, again, keep in mind that the composition of super-Earths is still a topic of debate, particularly whether they can have important amounts of volatile elements. Moreover, as explained in the Introduction, it is possible that a star that does form rocky planets is not able to retain the signature because those planets form when the star's convective envelope is too massive. Thus, the absence of a $\Delta\xfe$ versus $\tc$ trend consistent with the expectation for one particular rocky planet-host does not necessarily invalidate M09's hypothesis. 

GH13's work and ours demonstrate that in order to solve the apparent discrepancies and contradictions discussed above, the statistics of low-mass planets around Sun-like stars needs to improve significantly. We are currently contributing to these efforts by looking for planets around an important number of solar twin stars using ESO's HARPS spectrograph. All these objects will also be subject to a high-precision chemical abundance analysis, for which extremely high-quality data have been acquired using the MIKE spectrograph on the 6.5\,m Clay Magellan Telescope. With these data we will be able to provide a clearer picture of chemical signatures of planet formation in forthcoming publications.

\subsection{Ionization potential, age, and activity effects}

It is well known that condensation temperature correlates with the elements' first ionization potential \cite[FIP; see, e.g., Figure~5 in][]{ramirez10}. \cite{ramirez10} showed that for solar twins the statistical significance of the $\tc$ trend is higher than that of the FIP trend. Given the larger size of our sample and the high-quality of our analysis, we can now re-evaluate the FIP trends and attempt to find an alternative explanation for our observations. As in our previous work, we employed ionization potential values listed in the ``Atomic Properties of the Elements'' compilation by the National Institute of Standards and Technology (NIST).\footnote{Available online at http://www.nist.gov/physlab/data/periodic.cfm} First, we computed $\Delta\xfe$ versus FIP slopes in an identical manner as was done for $\tc$, i.e., using the mean abundance ratios of all stars as reference. Then, we defined a sample of ``pristine composition'' stars based on one known physical mechanism that, albeit unlikely, could be affecting the observed abundances, namely the so-called FIP effect.

Although the FIP trends may be attributed to small residual systematic errors in our model atmosphere analysis (differential non-LTE effects, for example), there is no clear picture of the actual physics responsible for the observed correlations. This makes it impossible to tell which stars are more sensitive to those alleged effects and to determine which stars have unaltered abundances. In other words, attributing the trends to unknown systematic errors would prevent us from investigating the problem further. On the other hand, it is tempting to attribute our observations to one related effect observed in the Sun, namely the so-called FIP effect. As shown by \cite{feldman92}, the coronal abundances of low FIP ions are about four times greater than those seen in the photosphere. \cite{henoux98} suggest that this is due to the acceleration of low FIP ions from the lower atmosphere by magnetic fields. If these ions come from the photosphere, which is actually not expected to be the case, then over time low FIP ions will be depleted. Thus, in this, albeit unlikely scenario, we can define a sample of stars with pristine composition by looking for objects that are not depleted in low FIP ions. Similar to the $\tc$ case, we define the pristine composition sample as the 15\,\% of stars with the most negative $\Delta\xfe$ versus FIP slopes (i.e., those with the highest low-FIP ion content), where the abundances of each star are measured relative to the average of all others. Relative abundance ratios were then recomputed using as reference the average abundances of the pristine sample. Hereafter we use these recomputed abundances.

FIP and $\tc$ are anti-correlated, which means that low FIP ions are generally refractory (high $\tc$) elements. Thus, pristine composition stars by definition have zero $\Delta\xfe$ versus FIP slope while the other stars have all positive FIP slopes because their low FIP abundances are smaller compared to the abundances of high FIP elements.

As in the $\tc$ case, the observed FIP slope distribution is wider and shallower than a pure noise distribution (cf.~Figure~\ref{f:slopes}), indicating that there is a real low FIP elemental abundance deficiency in some stars. The slope distributions of known planet hosts and other stars are not obviously different (cf.~Figure~\ref{f:slope_ph}), indicating that the presence of large planets cannot explain the FIP trends either.

The only potentially reasonable explanation for the low FIP elemental abundance deficiency could be that stellar winds take more of those elements away from the star. Such effect would be age dependent. Thus, we examined both the $\tc$ and FIP trends as a function of stellar age.

\begin{figure}
\centering
\includegraphics[bb=85 370 450 770,width=9.3cm]
{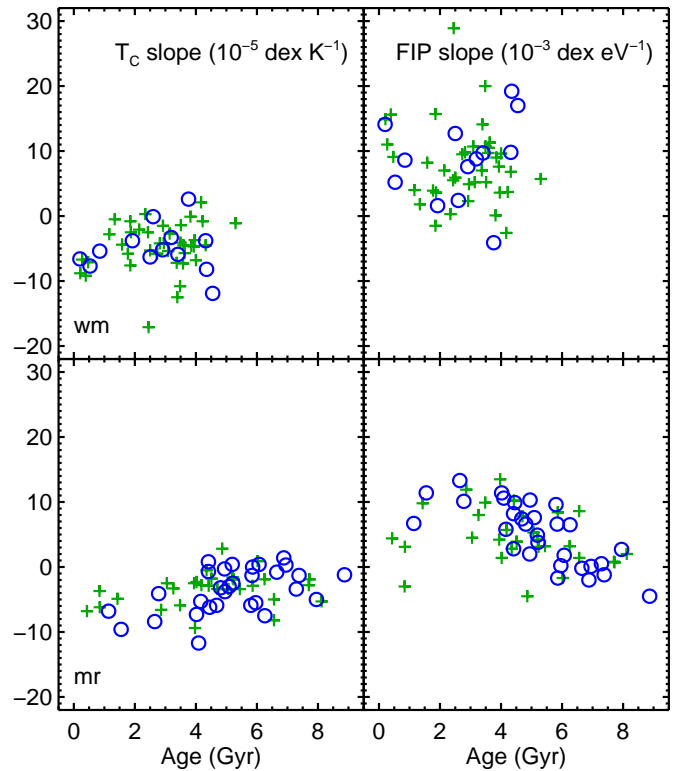}
\caption{$\Delta\xfe$ versus $\tc$ (left panels) and FIP (right panels) slopes as a function of stellar age for our warm (top panels) and metal-rich (bottom panels) samples. Known planet hosts are shown with open circles.}
\label{f:slopes_age}
\end{figure}

Interestingly, Figure~\ref{f:slopes_age} shows that there are important correlations between the slopes and stellar age, in particular for the metal-rich solar analogs sample. The $\tc$ correlation is weaker for the warm F-dwarf sample, and possibly non-existent for the FIP case in that group of stars. In any case, the most refractory-element depleted stars seem to be younger than the pristine composition objects. Alternatively, low FIP ions are more deficient in the younger stars. The latter contradicts the only known possible physical explanation for these trends, namely that low FIP ions are carried away from the star by its wind. In that scenario, we would expect those elements to be most depleted in older stars, which is clearly not the case.

It is not possible to explain in the context of our planet signature hypothesis why there are no young pristine composition stars. The signature is expected to be imprinted within the first few million years of the stars' lives. In its simple form, the hypothesis implies that a fraction of stars are not affected by this process, and thus even at 1--3 Gyr they could have pristine composition, but we observe none of those objects. A tempting idea could be that stars recover their lost refractories over time by accreting planet-like material.

\begin{figure}
\centering
\includegraphics[bb=85 370 450 770,width=9.3cm]
{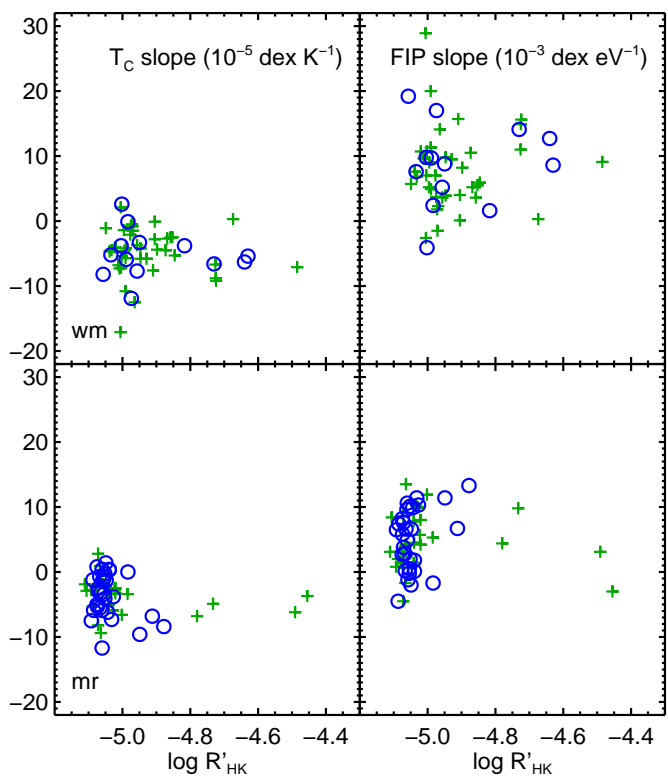}
\caption{$\Delta\xfe$ versus $\tc$ (left panels) and FIP (right panels) slopes as a function of the chromospheric activity index $\log R'_\mathrm{HK}$ for our warm (top panels) and metal-rich (bottom panels) samples. Known planet hosts are shown with open circles.}
\label{f:slopes_rhk}
\end{figure}

If Figure~\ref{f:slopes_age} is indicative of systematic errors in our analysis, the first place to look for additional evidence is in the stellar activity. The importance of the latter decreases with time and thus one would expect the older stars to have more reliable measured abundances. Figure~\ref{f:slopes_rhk} shows the relation between $\tc$ and FIP slopes and the chromospheric activity index $\log R'_\mathrm{HK}$. Nearly all our stars have low activity, which is in fact an observational bias imposed by our necessity to derive reliable abundances using models that do not take into account magnetic fields or starspots. There are no correlations between activity and the observed abundance slopes. If activity is the reason for the abundance trends then we would expect the most active stars in our samples to be on one of the extremes of the slope distribution, i.e., zero slopes or very negative (for the $\tc$ trends) or positive (for the FIP trends) slopes. Instead, they are distributed more or less randomly. Therefore, the dependency of slopes, either $\tc$ or FIP, with age cannot be physically attributed to an activity-related effect, making it unlikely that such systematic error in our model atmosphere analysis is responsible for the observed trends.

\section{Conclusions}

We have been able to detect small ($\lesssim0.15$\,dex) refractory-element depletions in stars other than solar twins by exploiting a purely differential approach. Instead of employing a solar spectrum as reference, our work is based on the determination of relative abundances of each star relative to all others in its group. Since all stars within each group are very similar to each other, the impact of systematic errors in the abundance analysis is minimized. By avoiding the use of a solar spectrum for reference, we also make negligible the impact of Galactic chemical evolution on the derived abundance trends, particularly the correlation between relative abundances and condensation temperature. 

The refractory element depletions that we observe are similar in amplitude to those detected in solar twin stars, allowing us to interpret them also as signatures of rocky planet formation. The depletion amplitudes are not strongly correlated with the presence of giant planets. Only a weak trend is detected after removing a potential $\teff$-dependence on the depletion amplitudes such that known planet hosts appear to be marginally more depleted in refractory elements. However, the observed difference is unlikely to account for the proposed depletion due to the formation of rocky material. This implies that, although more refractory element depletion may be observed when gas giant planets are present, their formation alone may not be able to explain the full effect.

We set out to test whether the amplitude of the chemical signature of terrestrial planet formation was dependent on the size of the stars' convective envelopes. Thus we examined abundance trends of stars that have significantly thinner convective envelopes compared to solar twin stars. Although we do observe that the maximum amplitude of refractory-element depletion appear to increase with higher $\teff$, it does so only up to about 5900\,K. Warmer stars present a nearly constant maximum depletion amplitude. Moreover, the maximum amplitudes of refractory-element depletion are very similar between our warm stars, metal-rich solar analogs, and the previously studied solar twin stars.

This could be explained in a number of ways. It could be that when rocky material forms around these three types of stars, much more of that material is formed in solar twins than metal-rich solar analogs, and not very much around the warm stars. Alternatively, it could be that similar amounts of rocky material are in fact formed around the three types of stars, and that the amplitude of depletions that we observe is regulated by the convective envelope size at the time those planets form. Based on our results, the convective envelopes of solar twin stars would have to be very thin, in fact at their final main-sequence sizes, when these planets form. In metal-rich solar analogs, the convective envelopes would need to be slightly more massive in order to dilute the chemical signature of a similar amount of rocky material. Finally, warm stars need even more massive convective envelopes when rocky planets form around them. This could be interpreted as planets being formed late in solar twins and early in warm stars because the convective envelope size decreases with time. Indeed, disk lifetimes appear to be shorter in more massive stars, in agreement with this interpretation. Furthermore, according to classical models of early stellar evolution, the convective envelopes of more massive stars reach their final sizes quicker than less massive stars, strengthening the effect suggested before. Episodic accretion will certainly complicate this picture, and add an element of randomness in the form of an accretion rate history that varies from star to star.

Due to the correlation between condensation temperature and first ionization potential (FIP), we have investigated trends of element depletions with FIP. The one unlikely, but known mechanism that could be affecting the observed compositions, namely the FIP effect, cannot explain our data because the trends do not follow the expected age dependency. On the other hand, the anticorrelation that we find between refractory element depletion and age may be a signature of stars recovering planet-like material over their main-sequence lifetimes. Finally, if FIP trends are due to systematic errors, we have found that they cannot be fully attributed to stellar activity.

Given the many variables involved, the results presented in this paper cannot conclusively confirm or reject the hypothesis that the formation of rocky material leaves a detectable signature on the stars' photospheric chemical compositions. Nevertheless, our work provides important additional clues that will help putting together a fully consistent picture in a near future. Certainly, larger samples of stars, including solar twins, warm late F-type dwarfs, metal-rich solar analogs, and others will help tackling the problem in a more statistically significant manner. A complete census of their planet populations would be ideal, but clearly unrealistic at this point. We are carrying out a number of efforts that aim at improving our knowledge of this particular field of exoplanet research.

\begin{acknowledgements}
We thank the anonymous referee for helping us improve our manuscript. I.R.'s work was performed under contract with the California Institute of Technology (Caltech) funded by NASA through the Sagan Fellowship Program. J.M.\ thanks FAPESP (2012/24392-2). M.A.\ gratefully acknowledges generous support from Australian Research Council (grants FL110100012 and DP120100991). This research has made use of the Exoplanet Orbit Database and the Exoplanet Data Explorer at exoplanets.org.
\end{acknowledgements}


\input{ms.bbl}
\end{document}

%% file: sample_small.tex
\\ \multicolumn{7}{c}{wm: late F dwarfs} \\ \hline
   522 & 5.7 & 6260 & 4.32 &  0.06 &  8 &   1.306 \\
  3119 & 7.4 & 6209 & 4.28 &  0.09 &  1 &     --- \\
  3236 & 6.5 & 6223 & 4.26 &  0.08 &  1 &     --- \\
  3540 & 7.0 & 6149 & 4.28 &  0.02 &  1 &     --- \\
  5862 & 5.0 & 6118 & 4.34 &  0.16 & 11 &     --- \\
  5985 & 6.5 & 6076 & 4.30 &  0.10 &  2 &     --- \\
  7978 & 5.5 & 6138 & 4.45 & -0.03 &  7 &   0.925 \\
  8548 & 7.1 & 6070 & 4.29 &  0.05 &  3 &     --- \\
 12653 & 5.4 & 6173 & 4.49 &  0.19 & 13 &   2.047 \\
 12764 & 7.1 & 6206 & 4.32 &  0.08 &  1 &     --- \\
\vdots &  \vdots &  \vdots &  \vdots &  \vdots &  \vdots & 
\vdots \\
\\ \multicolumn{7}{c}{mr: metal-rich solar analogs} \\ \hline
  1499 & 6.5 & 5731 & 4.37 &  0.22 & 11 &   0.024 \\
  1803 & 6.4 & 5798 & 4.45 &  0.20 & 11 &     --- \\
  5176 & 8.1 & 5858 & 4.39 &  0.16 &  2 &     --- \\
 12048 & 6.8 & 5776 & 4.16 &  0.14 & 13 &   0.250 \\
 12186 & 5.8 & 5840 & 4.14 &  0.15 &  9 &   0.067 \\
 17054 & 8.6 & 5760 & 4.08 &  0.34 &  2 &   1.405 \\
 17960 & 7.5 & 5856 & 4.39 &  0.23 &  3 &   3.836 \\
 20723 & 7.8 & 5681 & 4.43 &  0.24 & 10 &   5.797 \\
 20741 & 8.1 & 5780 & 4.41 &  0.20 &  3 &     --- \\
 21923 & 7.1 & 5761 & 4.25 &  0.26 &  1 &     --- \\
\vdots &  \vdots &  \vdots &  \vdots &  \vdots &  \vdots & 
\vdots \\

%% file: pars_small.tex
\\ \multicolumn{11}{c}{wm: late F dwarfs} \\ \hline
   522 & 6281 & 4.252 &  0.113 & 1.48 &  131 & 25 &  0.102 & 0.029 & -0.087 & 0.012 \\
  3119 & 6246 & 4.374 &  0.127 & 1.30 &   96 & 18 &  0.224 & 0.032 & -0.073 & 0.011 \\
  3236 & 6309 & 4.399 &  0.097 & 1.25 &  159 & 25 &  0.249 & 0.036 & -0.103 & 0.014 \\
  3540 & 6104 & 4.186 &  0.002 & 1.12 &  -46 & 16 &  0.036 & 0.032 & -0.198 & 0.015 \\
  5862 & 6111 & 4.359 &  0.173 & 1.18 &  -39 & 12 &  0.209 & 0.032 & -0.027 & 0.012 \\
  5985 & 6106 & 4.414 &  0.084 & 1.19 &  -44 & 11 &  0.264 & 0.029 & -0.116 & 0.010 \\
  7978 & 6114 & 4.412 & -0.023 & 1.13 &  -36 & 11 &  0.262 & 0.036 & -0.223 & 0.014 \\
  8548 & 6043 & 4.277 &  0.016 & 1.22 & -107 & 11 &  0.127 & 0.029 & -0.184 & 0.012 \\
 12653 & 6147 & 4.419 &  0.156 & 1.16 &   -3 & 11 &  0.269 & 0.029 & -0.044 & 0.010 \\
 12764 & 6211 & 4.349 &  0.082 & 1.28 &   61 & 12 &  0.199 & 0.025 & -0.118 & 0.009 \\
\vdots &  \vdots &  \vdots &  \vdots &  \vdots &  \vdots &  \vdots &  \vdots & 
\vdots &  \vdots & 
\vdots \\
\\ \multicolumn{11}{c}{mr: metal-rich solar analogs} \\ \hline
  1499 & 5729 & 4.365 &  0.139 & 1.06 &  -21 &  9 &  0.065 & 0.024 & -0.111 & 0.013 \\
  1803 & 5837 & 4.516 &  0.178 & 1.31 &   87 & 16 &  0.216 & 0.024 & -0.072 & 0.016 \\
  5176 & 5864 & 4.385 &  0.135 & 1.10 &  114 & 16 &  0.085 & 0.024 & -0.115 & 0.012 \\
 12048 & 5769 & 4.164 &  0.105 & 1.17 &   19 & 14 & -0.136 & 0.024 & -0.145 & 0.016 \\
 12186 & 5817 & 4.146 &  0.096 & 1.22 &   67 & 15 & -0.154 & 0.024 & -0.154 & 0.015 \\
 17054 & 5826 & 4.249 &  0.346 & 1.24 &   76 & 10 & -0.051 & 0.017 &  0.096 & 0.010 \\
 17960 & 5804 & 4.267 &  0.159 & 1.14 &   54 & 11 & -0.033 & 0.020 & -0.091 & 0.012 \\
 20723 & 5638 & 4.356 &  0.196 & 0.96 & -112 &  9 &  0.056 & 0.020 & -0.054 & 0.011 \\
 20741 & 5826 & 4.519 &  0.145 & 1.19 &   76 & 16 &  0.219 & 0.024 & -0.105 & 0.017 \\
 21923 & 5784 & 4.197 &  0.220 & 1.17 &   34 & 11 & -0.103 & 0.017 & -0.030 & 0.012 \\
\vdots &  \vdots &  \vdots &  \vdots &  \vdots &  \vdots &  \vdots &  \vdots & 
\vdots &  \vdots & 
\vdots \\

%% file: ll_small.tex
 Fe I & 1334 & 5775.0801 & 4.220 & -1.300 \\
 Fe I &      & 5778.4531 & 2.588 & -3.440 \\
 Fe I &      & 5793.9141 & 4.220 & -1.619 \\
 Fe I &      & 5806.7300 & 4.610 & -0.950 \\
 Fe I &      & 5809.2178 & 3.883 & -1.710 \\
\vdots &  \vdots &  \vdots &  \vdots & 
\vdots \\
  V I & 1429 & 6039.7271 & 1.063 & -1.854 \\
      &      & 6039.7290 &       & -1.854 \\
      &      & 6039.7300 &       & -2.030 \\
      &      & 6039.7329 &       & -1.690 \\
      &      & 6039.7329 &       & -2.155 \\
      &      & 6039.7339 &       & -2.280 \\
      &      & 6039.7378 &       & -1.682 \\
      &      & 6039.7378 &       & -1.716 \\
      &      & 6039.7388 &       & -2.708 \\
      &      & 6039.7441 &       & -1.843 \\
      &      & 6039.7441 &       & -1.433 \\
      &      & 6039.7510 &       & -1.217 \\
  V I &      & 6081.4170 & 1.051 & -1.660 \\
      &      & 6081.4170 &       & -1.484 \\
      &      & 6081.4268 &       & -1.484 \\
      &      & 6081.4282 &       & -1.359 \\
      &      & 6081.4419 &       & -1.359 \\
      &      & 6081.4419 &       & -1.677 \\
      &      & 6081.4419 &       & -1.472 \\
      &      & 6081.4600 &       & -1.472 \\
\vdots &  \vdots &  \vdots &  \vdots & 
\vdots \\

%% file: abundances_small.tex
HIP & $\Delta$[C/H]  & $\sigma$ & $\Delta$[O/H]  & $\sigma$ & $\Delta$[Na/H] & $\sigma$ & \ldots \\ \hline
\\ \multicolumn{8}{c}{wm: late F-dwarfs} \\ \hline
   522 & -0.059 & 0.036 & -0.066 & 0.011 & -0.183 & 0.011  & \ldots \\
  3119 & -0.106 & 0.033 & -0.031 & 0.002 & -0.192 & 0.011  & \ldots \\
  3236 & -0.213 & 0.044 & -0.117 & 0.003 & -0.242 & 0.003  & \ldots \\
  3540 & -0.146 & 0.080 & -0.164 & 0.005 & -0.288 & 0.003  & \ldots \\
  5862 & -0.041 & 0.011 & -0.032 & 0.005 & -0.097 & 0.004  & \ldots \\
  5985 & -0.157 & 0.060 & -0.136 & 0.005 & -0.228 & 0.008  & \ldots \\
  7978 & -0.260 & 0.076 & -0.230 & 0.020 & -0.440 & 0.008  & \ldots \\
  8548 & -0.089 & 0.059 & -0.142 & 0.006 & -0.236 & 0.026  & \ldots \\
 12653 & -0.106 & 0.036 & -0.032 & 0.012 & -0.179 & 0.030  & \ldots \\
 12764 & -0.076 & 0.041 & -0.069 & 0.010 & -0.229 & 0.011  & \ldots \\
\vdots & \vdots & \vdots & \vdots & \vdots & \vdots & \vdots &
 \vdots \\
\\ \multicolumn{8}{c}{mr: metal-rich solar analogs} \\ \hline
  1499 & -0.103 & 0.023 & -0.099 & 0.015 & -0.134 & 0.011  & \ldots \\
  1803 & -0.215 & 0.031 & -0.086 & 0.021 & -0.206 & 0.015  & \ldots \\
  5176 & -0.130 & 0.030 & -0.121 & 0.015 & -0.148 & 0.004  & \ldots \\
 12048 & -0.229 & 0.049 & -0.141 & 0.012 & -0.340 & 0.014  & \ldots \\
 12186 & -0.253 & 0.029 & -0.155 & 0.008 & -0.298 & 0.031  & \ldots \\
 17054 &  0.156 & 0.034 &  0.108 & 0.022 &  0.234 & 0.013  & \ldots \\
 17960 & -0.126 & 0.036 & -0.080 & 0.009 & -0.183 & 0.007  & \ldots \\
 20723 & -0.027 & 0.020 & -0.031 & 0.023 & -0.059 & 0.026  & \ldots \\
 20741 & -0.244 & 0.032 & -0.102 & 0.028 & -0.263 & 0.008  & \ldots \\
 21923 & -0.153 & 0.032 & -0.065 & 0.009 & -0.177 & 0.011  & \ldots \\
\vdots & \vdots & \vdots & \vdots & \vdots & \vdots & \vdots &
 \vdots \\

%% file: agerhk_small.tex
\\ \multicolumn{8}{c}{wm: late F dwarfs} \\ \hline
   522 & 1.26 & 1.22 & 1.29 & 2.60 & 1.95 & 2.99 & -4.98 \\
  3119 & 1.22 & 1.20 & 1.24 & 1.16 & 0.17 & 2.16 & -4.91 \\
  3236 & 1.22 & 1.21 & 1.25 & 0.39 & 0.12 & 1.87 & -4.72 \\
  3540 & 1.22 & 1.16 & 1.26 & 3.82 & 3.67 & 5.01 & -4.91 \\
  5862 & 1.18 & 1.17 & 1.20 & 2.14 & 0.67 & 2.93 & -4.98 \\
  5985 & 1.14 & 1.13 & 1.16 & 1.58 & 0.18 & 2.75 & -4.90 \\
  7978 & 1.12 & 1.11 & 1.14 & 1.92 & 0.35 & 3.25 & -4.82 \\
  8548 & 1.13 & 1.11 & 1.17 & 4.17 & 3.70 & 4.50 & -5.01 \\
 12653 & 1.19 & 1.18 & 1.20 & 0.85 & 0.31 & 1.79 & -4.63 \\
 12764 & 1.19 & 1.18 & 1.20 & 1.85 & 1.22 & 2.70 & -4.97 \\
\vdots & \vdots & \vdots & \vdots & \vdots & \vdots & \vdots &
 \vdots \\
\\ \multicolumn{8}{c}{mr: metal-rich solar analogs} \\ \hline
  1499 & 1.03 & 1.02 & 1.05 & 6.07 & 4.62 & 7.05 & -5.04 \\
  1803 & 1.09 & 1.08 & 1.11 & 0.85 & 0.06 & 1.97 & -4.49 \\
  5176 & 1.08 & 1.06 & 1.10 & 3.94 & 2.61 & 4.77 & -5.02 \\
 12048 & 1.06 & 1.04 & 1.09 & 7.95 & 7.41 & 8.43 & -5.07 \\
 12186 & 1.08 & 1.06 & 1.11 & 7.29 & 6.76 & 7.77 & -5.05 \\
 17054 & 1.20 & 1.16 & 1.21 & 4.41 & 4.12 & 4.53 & -5.07 \\
 17960 & 1.09 & 1.07 & 1.12 & 5.96 & 5.24 & 6.92 & -5.07 \\
 20723 & 1.02 & 1.01 & 1.04 & 6.88 & 5.78 & 7.90 & -5.05 \\
 20741 & 1.08 & 1.06 & 1.10 & 0.84 & 0.04 & 2.04 & -4.45 \\
 21923 & 1.16 & 1.09 & 1.18 & 6.56 & 5.13 & 7.12 & -5.07 \\
\vdots & \vdots & \vdots & \vdots & \vdots & \vdots & \vdots &
 \vdots \\